# Evaluating the importance of environmental factors on the spatial distribution of livestock settlements in the Monte desert with a Monte Carlo based model: Settlement Dynamics in Drylands (SeDD).

Emmanuel N. Millán[a,b], Silvana Goirán[a,c], Julieta N. Aranibar[a,c,*], Leonardo Forconesi, Carlos García Garino, Eduardo M. Bringa[a]

[a]Facultad de Ciencias Exactas y Naturales, Universidad Nacional de Cuyo. Padre Contreras 1300. Parque Gral. San Martín. Mendoza, M5502JMA. Argentina.
[b]Institute for Information and Communication Technologies (ITIC), Universidad Nacional de Cuyo, Parque Gral. San Martín. Mendoza, 5500, Argentina.
[c]Instituto Argentino de Nivología, Glaciología y Ciencias Ambientales (IANIGLA), CONICET, CCT-Mendoza. Ruiz Leal s/n, Parque Gral. San Martín, CC 330. Mendoza, 5000. Argentina
*Corresponding author: jaranibar at mendoza-conicet.gob.ar

**Abstract**

In the Monte desert, increasing population density, changing land rights and infrastructure may encourage livestock activity, with unknown consequences on ecosystems. Factors that influence livestock settlement distribution may affect ecosystem degradation. We hypothesize that surface and groundwater availability influence livestock settlements distribution. We evaluated this hypothesis with a Monte Carlo based model to simulate Settlement Dynamics in Drylands (SeDD), which calculates probabilities on a gridded region based on six environmental factors: groundwater depth, vegetation type, proximity to rivers, paved roads, old river beds, and existing settlements. A parameter sweep, including millions of simulations, was run with combinations of parameters related to these factors. The sets of parameter values that minimized the residuals between simulations and observations indicated the relative importance of each factors on settlement distribution. Distances to rivers and old river beds were critical to explain the current distribution of settlements, while vegetation, paved roads, and water table depths were not important. Spatial distribution of simulated vegetation, which included degradation around livestock settlements, generally agreed with remotely sensed vegetation classes. The model could be a useful tool to evaluate the effects of land use changes (water provision, river flows), on settlement distribution and vegetation degradation in arid environments.

**1.Introduction**

Livestock production, the largest land use sector on Earth, is experiencing changes related to climate change and anthropogenic pressures (Schneider, 2010). Population and economic growth, urbanization, and consumption patterns are shaping livestock production, with impacts on societies and environments, such as greenhouse gas emissions, nutrient cycles, land demand and degradation, and protein supply (Herrero and Thornton, 2013). The challenge to feed the world sustainably partly depends on how we understand and manage the livestock sector. In drylands, which sustain a third of the world population and 78 % of livestock worldwide (Asner et al., 2004; Corvalan et al., 2005) livestock production is one of the main economic activities.

Groundwater coupled ecosystems in the Monte desert (Argentina) are used for subsistence livestock production, which allows the coexistence of areas with high vegetation cover in most of the region (Goirán et al., 2012), and rural communities. However, changing land rights (Gobierno de la Provincia de Mendoza, 2001), water provision, infrastructure, and population growth may increase population density and grazing pressures, with increasing risks of ecosystem degradation. In order to predict future conditions of livestock production and ecosystems in the region, it is crucial to understand the feedbacks between natural resources and livestock settlements at present.

Numerical models are useful to simulate, as in a laboratory, controlled experiments where initial conditions and system feedbacks are changed (Bankes, 2002; Bankes *et al.*, 2002). Environmental models can be used to simulate landscapes, and test hypotheses about human-environment interactions. With enough data to constrain parameters, these models can be useful tools for prediction and management of environmental resources. Growth, migration, and change of human settlements have been modeled for ancient societies and landscapes with different approaches. Agent-based models (ABM) have been used to simulate past societies, like the Anasazi (also called Pueblo). ABM simulate the behavior of individual "agents" based on environmental, social, and political factors, such as food and water availability, social status and organization (households, villages, tribes), marriage, and agricultural practices (Crabtree and Kohler, 2012; Kohler et al., 2012). Households were defined as agents, with attributes such as age, location, grain stocks, death age, and nutritional needs. The landscape for the simulation contained information on maize crop yield, surface and groundwater, soil type, soil degradation, wood, forage, and hunting animals (Axtell et al., 2002; Dean et al., 2000; Kohler et al., 2012). A similar approach, including soil erosion, was used to study prehistoric settlements around the world (Barton et al., 2010a and b). These models may simulate complex and numerous processes, but in cases where data to validate different components of model behavior are not available, it may be difficult to evaluate whether the added complexity contributes significantly to understand real ecosystems.

Other models, such as habitat models, simulate plant and animal habitat ranges with correlation and mechanistic approaches, based on known species distributions or physiological requirements (Kearney and Porter, 2009; Guisan and Zimmermann, 2000). Dispersion of plants has been simulated with spatially explicit, stochastic models, based on environmental conditions relevant for their survival (i.e., abiotic features and habitat type), and dispersion (i.e., corridors such as roads and rivers) (Fennell et al., 2012). Land cover changes in the Amazon, from mature forests to pastures, have been simulated with a model that assigns different land uses based on decision making rules that depend on demographic rates (fertility, migration), institutions, and agricultural prices (cattle and rice) (Evans et al., 2001). The aforementioned models have advantages and disadvantages for each case, depending on the balance between model complexity and data availability to constrain parameters and test results.

In arid environments, used for extensive livestock production, several environmental and socioeconomic drivers may influence the dynamics of livestock settlements. Availability of woodland resources, groundwater, and access roads may all affect settlement patterns, but their interactions or relative importance on settlements distribution and success are not known. Studies of present and historic human occupation in NE Mendoza report a heterogeneous distribution, with spatial aggregations around rivers and other landscape features (i.e., old river beds), indicating the importance of water availability for human settlements (Chiavazza, 1012 and 2014; Chiavazza and Prieto, 2008; Goirán et al., 2012). However, the relative importance of surface, groundwater, and old river beds is difficult to obtain from a simple observation of settlement distribution, because more than one factor may overlap (Gomez et al., 2014), and have opposing or multiplying effects on a given space. Goirán et al. (2012) also found a pattern of concentric vegetation reductions around settlements, given by the concentration of animals and higher pressures around water points, also observed in other deserts (Ringrose et al., 1996). The practice of night-time livestock accumulation, free grazing around settlements, and the scarcity of permanent fresh water sources generate concentric gradients of grazing pressure. Because environmental and economic changes may encourage or discourage settlement establishments and change their distribution in the landscape, we aim to identify the main drivers of landscape occupation in these arid groundwater-coupled ecosystems. We hypothesize that surface and groundwater availability are the most important factors for settlements in drylands, with a minor effect of woodland resources and access roads. In order to test this hypothesis for the Monte desert, we used a Monte Carlo based model of Settlement Dynamics in Drylands (SeDD), which includes six environmental drivers of settlements: surface and groundwater availability, vegetation type, existing settlements, access routes, and old river beds. These factors provide different services to settlers, such as water provision for humans and livestock, woodland products for construction and forage, transport and communication with existing settlements and other regions, and initial labor, materials, and water during the construction period. The model assumes that places with higher availability of water and woodland resources will be preferentially settled, independently of social structure. The model assigns settlements in places where the aptitude (simulated as probabilities) is higher, with a number of settlements established stochastically. The model also simulates the degradation of vegetation around settlements, gradually reducing the suitability of these spaces. Our model differs from plant and animal dispersion models (i.e., Fennell et al., 2012) because it assumes that settlers have a prior knowledge of the environment in the entire region, simulating environmentally-based human informed decisions. Millions of simulations with combinations of crucial parameters were run to find minimum residuals between simulated and observed spatial indexes of settlements in relation to other settlements, rivers, and roads. Combinations of parameter values that resulted in low residuals were interpreted to indicate the relative importance of each parameter.

## 2. Materials and Methods
### 2.1 Study area description.

Our study area is located in the non irrigated lands of North East (NE) Mendoza, Argentina, where mean annual precipitation is below 200 mm. The region is framed by permanent and temporary rivers (San Juan, 68° 1' S, Desaguadero, 32° 41'W, Mendoza, 64° 34'W, and Tunuyán, 33° 16'S, rivers) (Fig. 1). Groundwater is recharged in the Andes (200 km West) and reaches the area with a high salt and arsenic content, preventing its use for irrigation (Aranibar et al., 2011). The region has an aeolian plain with sand dune-interdune systems, old river beds, and lacustrine systems (Fig. 1), with varying access to surface and groundwater. Most of the region is occupied by the aeolian plain, and lacks surface water. One of the four old river beds of the region crosses the aeolian plain from West to East, providing an easier access to the territory, and localized patches of groundwater with a better quality (Jobbagy et al., 2011; Aranibar et al., 2011). The other old river beds are shorter, and interrupted by sand dunes. Historic documents suggest that river beds have been dried at least from 1778 (Prieto, 2000). The only paved road of the region (road Nº 142) was built along the main old river bed for most of its length. People live in livestock settlements, which mostly hold 1 to 3 families and their livestock (mainly goats, but also cattle and horses). The housing, corral and well structures, where livestock accumulate and shrubs are removed, are generally located in circular to oval areas of 50 to 100 m in diameter (supporting kml map). At present, there are 577 settlements with a heterogeneous spatial distribution, aggregated at different scales: at a regional scale, settlement densities are higher near rivers, while at a local scale, settlements are generally close to other settlements (Goirán et al., 2012). Settlements located far from the paved road are accessed through dirt roads that cross the high sand dunes of the aeolian plain, decreasing the possibilities of communication, trade, and transportation between areas (Goirán et al., 2013).

In interdune valleys where groundwater is near the surface (from 5 to 15 m depth), highly productive, phreatophyte, *Prosopis flexuosa* woodlands develop (Contreras et al., 2010; Jobbagy et al., 2011). These woodlands have been seasonally used by aboriginal groups since pre-hispanic times, providing them with hunting animals and *Prosopis* pods (García Llorca and Cahiza, 2007). During colonial times, many aboriginal individuals or groups used the area as a refuge, changing the previous seasonal and complementary occupation of the area to a more permanent pattern of occupation (Escolar, 2007; Prieto, 2000). During the 19[th] century, woody species, particularly *P. flexuosa, Bulnesia* *retama* and *Capparis atamisquea*, were cut for railroad and vineyard construction in irrigated oases, and domestic fuel in developing cities, but sand dunes prevented clear cutting in certain areas, where old *P. flexuosa* individuals still remain (Alvarez et al., 2006; Rojas et al., 2009; Villagra et al., 2005).

At present, local Huarpe descendants still inhabit these lands, mainly practicing subsistence livestock production in permanent livestock settlements, which rely exclusively on groundwater for most uses. Animals graze freely during the day around the settlement, but return at the end of the day to drink water, and they are kept in corrals during the night. In areas close to paved roads, people have access to drinking water, transported from irrigated oases by trucks, or a recently (2010) built aqueduct. The hydrogeology of the region, including a shallow aquifer and fine sediments (Aranibar et al., 2011; Gomez et al., 2014), allows the construction of wells by independent individuals at a relatively low cost, without government assistance. Wells are constructed near the corral and housing area with wooden *Prosopis flexuosa* frames, and groundwater is extracted manually, or with the help of an animal. This relative independence of livestock owners from government assistance and planning allows settlers to establish in areas that they consider appropriate for their subsistence, probably basing their decisions on their knowledge of natural resources availability, as we simulate with our model.

The exclusive reliance of livestock on groundwater from their settlements causes a pattern of night-time animal concentration around wells and corrals, as also observed in Botswana, Patagonia and other arid areas (Ringrose et al., 1996; Bisigato and Bertiller, 1997). This causes higher pressures near wells, and consequent changes on soils, groundwater quality, and vegetation (Goirán et al., 2012; Aranibar et al., 2011; Meglioli et al., 2013). Most changes are visible in rings of 2 km around livestock settlements, although overlap of grazing areas with high settlement densities may cause vegetation cover and biomass reductions in larger areas (Goirán et al., 2012; Ringrose et al., 1996).

**2.2 Model description for the Monte desert.**

We modeled settlement dynamics to test the hypothesis that natural resources, mainly water, play a determinant role in the decision to establish new livestock settlements, and their success as productive units. For this purpose, we developed a Monte Carlo based model, which is presented in detail in the Supporting Online Material (SOM). We simulate the aptitude or suitability of the landscape as probabilities for settlement, based on the partial probabilities given by different environmental conditions. We assumed that local inhabitants have a knowledge of certain environmental features in the whole region, so they are capable of settling in any place of the region that they consider appropriate (i.e., the model selects, at each time step, the pixel with maximum probability found in the whole area). Then, areas with a higher suitability will be preferentially settled. The model also has a stochastic component to assign a proportion of the settlements in unfavorable places at random, controlled by the parameters *Pset* and *Pthresh* (Table 1 and SOM). This random component aims to incorporate unknown factors that may influence human decisions. We considered that rivers have a positive influence to settlements up to 5 km away, because animals can travel this distance to water sources. Shallow groundwater is the main water source for most settlements (including those close to rivers), which is extracted with manually built wells. For this reason, we include water table depth as a controlling factor for settlements.

Old river beds may also be preferential areas for settlement, because their smooth slopes and finer sediments than their surroundings allow rain water accumulation in soil ditches (jagüels), providing an additional, although temporary, freshwater source for domestic animals. Patches of groundwater with a better quality are found in old river beds (Aranibar et al., 2011; Gomez et al., 2014), and the smooth and uniform slopes could provide an easier access and better communication to other areas, probably favoring settlement dispersion.

Roads provide clear advantages to settlers, including a better access to commercial centers and irrigated agricultural areas, where people often work during the harvest season (Goirán et al., 2013). Fresh water is also transported from the oases in trucks, but they only reach accessible areas near the paved road. We assume that the road has a positive influence for settlements up to 5 km.

Vegetation provides most of the products needed for livestock activities. We classified the vegetation in different classes according to the dominant species and the general products that they provide: woodlands, shrublands, *chañarales (Geoffroea decorticans* small woodlands), and old river bed and floodplains vegetation. Woodlands are given the highest probability, because they offer wood for fuel and construction, and forage, and *P. flexuosa* pods for animal food.

Finally, during the construction period of a new settlement, a "mother" settlement provides water, tools, and labor for construction, until a well is functional (Torres, 2008), so we assumed that the aptitude of a pixel decreases with increasing distances from existing settlements.

In the following subsections, input grids, parameters and variables included in the model are described.

### 2.2.1. Model functioning:

The SeDD model divides the region of interest into a regular grid, made of square pixels, and estimates the "probability" of each pixel for being settled. Probabilities are given by the combination of six factors, assumed to affect settlement establishment and functioning in arid areas: groundwater depth, minimum distance to roads, rivers, and existing settlements, presence of old river beds, and pixel vegetation type. These factors are represented in the model in six different layers, with each pixel having a partial probability value for each factor, given by its characteristics and potential services to settlers. Total probabilities for each pixel are calculated by multiplying partial probabilities given by each of the six environmental factors, assuming that they act independently. The model then assigns a new settlement in the pixel with the maximum probability if a threshold probability value is surpassed, or at random if the threshold is not reached anywhere in the grid. After a settlement is established, the partial probabilities given by "vegetation type" in surrounding $24^{th}$ neighboring pixels gradually decrease, representing vegetation consumption by the livestock of a new settlement. Because of the stochastic nature of the model (i.e., a proportions of settlements established at random), multiple simulations are run for a given set of parameters or initial conditions, and simulation results are then averaged.

A flowchart showing how the simulations work is presented in SOM-Fig.1., and explained below:

a) The model reads the input files and parameters (rivers, roads, etc). Partial probabilities are evaluated in each pixel for each environmental factor, based on initial values of input grids and parameters. After each time step, these values are updated, including new settlements, and partial probabilities are re-calculated.

b) The probability associated with vegetation type (*PVeg*) is decreased around existing settlements in each time step.

c) The model calculates the total probabilities (*P*) for each pixel in the grid, by multiplying partial probabilities in each pixel. All pairs with maximum *P*, *P= Pmax*, are saved. If there is more than one pixel with the same maximum *P*, the model randomly chooses one of them. Partial probabilities for vegetation type and distance from existing settlements are recalculated at each time step, considering newly established settlements.

d) If *Pmax > Pset* (threshold of total probability *P* to place a settlement by high probability), the model assigns a new settlement in the pixel with maximum *P*.

e) If *Pmax<Pset*, the model may still place a random settlement. First a random number is used to select a single pixel from the grid and, then, a second random number from a uniform distribution in (0,1) is used to decide if a settlement will be established in that pixel. This decision is based on a pre-set rejection threshold (*Pthresh*). For instance, for a rejection rate of 70%, there will be a settlement in the randomly chosen pixel only if the second generated random number is greater than *Pthresh*=0.7. The rejection threshold is related to pressures which force settlements in places where environmental conditions are highly unfavorable.

f) Steps (a)-(e) are repeated for each time step until the end of the simulated time is reached, i.e. by running a number *TotalSteps* of iterations. Then, the model calculates pair correlation functions for the distance between settlements and rivers, settlements and road, and between existing settlements.

g) The user may transform the resulting grid of final *PVeg* values into a final vegetation map, assigning vegetation type classes to given ranges of *PVeg* values, according to Table 2.

**2.2.2 Input grids**

In order to simulate the area of interest, a square area of the region, which included the study area, was divided in 22500 (150 x 150) square pixels of 56.25 ha (750 m x 750 m) each. This resolution allows us to obtain pixels with and without a settlement, and with different disturbance intensities. At the same time, with this pixel size we can simulate vegetation degradation up to 2 km from the settlement, as observed in vegetation studies, including up to $5^{th}$ neighbors. A finer resolution would not allow us to simulate neighbor effects up to 2 km, because it would take a higher number of neighbors, and exceed our current computational capacity. Seven input grids were elaborated for the region using different maps.

a) *Initial settlements*. The layer of initial settlements to start the simulations was made using a map of settlements from 1928 (IGM, 1928), which was digitalized and transformed to a 750 m pixel raster.

b) *Paved road 142*. This layer was digitalized from a mosaic of orthorectified TM Landsat images of the study area (path 231 rows 082 and 083) acquired on March $8^{th}$ 2011, and transformed to 750 m pixel raster map. Because pixel size was chosen to optimize resolution and computational capacity to simulate processes in neighboring pixels, the resulting road map has a wider road than in the real terrain (750 m). However, with this resolution we are able to include all settlements located near the road. The road 142 was built in 1975, so we include the input road layer at the middle of the simulated period.

c) *River layer*. The San Juan, Desaguadero and Mendoza Rivers were digitalized from a mosaic of orthorectified TM Landsat images of the study area (path 231 rows 082 and 083) acquired on March $8^{th}$ 2011, and transformed to a 750 m pixel map.

d) *Initial Vegetation layer*. We made a map of the presumable vegetation of the area one hundred years ago, based on a map of the potential distribution of *P. flexuosa* woodlands, combined with current vegetation maps. The potential distribution of *P. flexuosa* woodlands was modeled with the Maxent software, using georeferenced data of the presence of *P. flexuosa* woodlands and relevant environmental factors (e.g., climatic, altitude, slope, aspect, soil and water table depth) (Perosa et al., 2014). We made the current vegetation map with a non supervised classification from a georeferenced Landsat 5 TM mosaic (path 231, rows 082 and 083) acquired in March $8^{th}$ 2011, using a cluster analysis approach with the ISODATA function on ENVI software (ENVI, 2003). 15 classes were obtained. The classes were regrouped into five different vegetation classes: woodlands, shrublands with high and low cover, *chañarales* (*Geoffroea decorticans*) and old river bed-floodplain vegetation. The vegetation type of each class was assigned by visual interpretation of Google Earth (Google, Mountain View, CA) high-definition images, and field observations. Comparing this classification with 50 field surveys (10 for each vegetation class) obtained in April and May 2011, 96% of field classification agreed with the vegetation map classification: 90% in woodlands, 90% in each, high and low cover shrublands, and 100 % in the other vegetation types (*chañarales* and old river bed-floodplain vegetation).Then the 30 m pixel image was resampled at 750 m pixel with a nearest neighbor resampling method (ENVI Resize Data Spatial/spectral module; ENVI, 2003), which uses the nearest pixel value as the output pixel value. There is not a historic vegetation map for the region, but the main changes may have occurred by clear cutting for railroad and vineyard construction, and animal foraging around settlements (Goirán et al., 2012; Alvarez et al., 2006). Based on this knowledge, we made a historical map transforming the current degraded areas (low cover shrublands) into high cover shrublands, and redefining woodlands to match the map of potential *P. flexuosa* woodlands elaborated from environmental factors (Perosa et al., 2014). The other vegetation classes, associated with old and present river beds (*chañarales* and floodplain vegetation) were not changed for the historic map. The resulting historic map has four vegetation classes: *Chañarales*, woodlandss, high cover shrublands, and old river bed-floodplain vegetation.

  e) *Water table depth*. Values for each pixel were calculated from a digital elevation model and a potentiometric map of the unconfined aquifer. The potentiometric map (indicating hydraulic heads, or water table heights above sea level, *wh*) was developed by Gomez et al. (2014), with a scale of study of 1 well every 150 km$^2$, yielding equipotential lines every 10 m. They measured groundwater depths of 138 wells in the unconfined aquifer, obtaining well depths between 5 to 120 m, which were interpolated with Kriging regressions. A value of hydraulic head was assigned to each 750 m pixel with interpolation. The surface elevation above sea level, *z*, was obtained from a digital elevation model (SRTM-DEM, Shuttle Radar Topography Mission -Digital Elevation Model) developed from radar data collected during the 2000 (USGS, 2004), and validated with local geodesic studies (Lenzano and Robin, 1995). The SRTM-DEM elevation data was 4.8 ± 1.33 m (mean and standard deviation, corresponding to accuracy and error, respectively) higher than the geodesic elevations, with a coefficient of determination of 0.997 (r$^2$) between the two data sets (Aranibar et al., 2011). Source for these data were the Global Land Cover Facility [http://www.landcover.org]. Values of hydraulic head were resampled at 750 m pixel map with a nearest neighbor resampling method (ENVI Resize Data Spatial/spectral module; ENVI, 2003). Elevation data were resampled to a 750 m pixel map. Water table depths for each pixel, *w*, result from the difference between elevation, *z*, and *wh*. This map was classified into 3 groundwater depth classes: less than 15 m (accessed by manual tools), 15-25 m (accessed with simple mechanical tools), more than 25 m (accessed with more complex and expensive, generally government-supported technology). These values are kept constant during the simulation, because low local precipitation does not cause significant recharges and *wh* fluctuations (Jobbagy et al., 2011).

  f) *Old river bed*. This grid was made from a geomorphological map of the area (Goirán et al., 2012; Gomez et al., 2014), which differentiates the old river beds from other geomorphological units.

  g) *Mask.* Pixels that fell outside the area of interest because they do not represent traditional livestock settlements (i.e., irrigated oases, mountains) were removed from the simulations using a mask of zero total probability values for settlements. This resulted in an irregular simulated region of 17465 pixels, framed by the rivers and an irrigated oasis. This treatment of boundary conditions is justified in the study area because rivers prevent animal movement and settlement interactions outside the defined area. Furthermore, outside the area, traditional settlements are uncommon because geopolitical or environmental factors, such as land and water rights, mountains, or taller dunes, favor other land uses (i.e., agriculture, unoccupied lands).

**2.2.3. Input parameters**

Model parameters are related to the partial probabilities of each pixel given by the hypothesized controlling factors: probabilities of pixels located at different distances from rivers, roads, and existing settlements, or pixels located on river beds and having different vegetation types. Other parameters are related to the proportion of settlements established at random or by high probability, and the dynamics of the model. A list of parameters is presented in SOM, and the values for the real case simulation are presented in Table 1. For the chosen time step of 0.6 month (20 time steps per year), the number of steps to run (2000 steps) is related to the amount of years (100 years) we wanted to simulate.

*Partial probabilities* for each environmental factor were calculated from the input grids, assigning probability values according to the following rules: best pixels (closer to relevant environmental features: roads, rivers, settlements, groundwater, old river bed, woodlands) have the highest probabilities ($P=1$); worst pixels (far from relevant environmental features) have the lowest probability ($P=0.1$); intermediate pixels have intermediate probabilities, whose relative importance were estimated with a parameter sweep.

a) *PRiverDist* and *PRoadDist:* In order to represent the decreasing positive influence of rivers and roads (given by access to water and communication) with increasing distances, we categorized *RiverDist* and *RoadDist* for each pixel in three possible classes: up to 2.9 km (between *RiverSqrmin* to *RiverSqrmed*, and *RoadSqrmin* to *RoadSqrmed*), 2.9 to 5.3 km (*RiverSqrmed* to *RiverSqrmax*, and *RoadSqrmed* to *RoadSqrmax*), and more than 5.3 km (higher than *RiverSqrmax* and *RoadSqrmax*). These distance classes relate to the daily walking effort of livestock and people, which is assumed to be minimum at less than 3 km. Because the road Nº 142 was built in 1975 (Dirección Nacional de Vialidad, personal communication), the road is included in the simulations from time step 1000. The maximum probabilities associated with rivers and roads were assigned to the first distance class (*RiverSqrmin* to *RiverSqrmed* and *RoadSqrmin* to *RoadSqrmed*), where *Privermax* =1 and *Proadmax* =1. Probabilities corresponding to the second distance class for road and rivers (*Privermed* and *Proadmed*) were selected with a parameter sweep, as described below. Probabilities of the third distance class were assumed to be 0.1 for *Proadmin*, while a parameter sweep was used to select *Privermin*.

b) *PSettDist:* The different distance classes between settlements were chosen based on the need to subsidize a new settlement with water and food during the construction period, until a new well is built. *Psettlmax* = 1 if the distance of a potential site to an existing settlement was smaller than 1.7 km and *Psettlmin* = 0.1 if the distance was larger.

c) *Probabilities given by vegetation type, PVeg.* Initial vegetation was classified into 4 classes: woodlands, high cover shrublands, *chañarales*, and old river bed-floodplain vegetation. Initial probabilities related to each vegetation type (*PVeg1* to *PVeg4*) (see Table 1) were chosen based on the resources offered by each vegetation type, such as forage and wood production, and palatable species. Woodlands, where *PVeg1*=1, are the most valuable class because they provide wood for multiple uses, such as construction material for fences, houses and wells, and animal forage during the entire year, including annual grasses during the summer, perennial shrubs, and *Prosopis* fruits, which are collected and stored for winter reserves (Allegretti et al., 2005). Shrublands provide forage during the entire year, but lower quality and quantity of wood, so the corresponding *Pveg2* value was assumed to be lower than *PVeg1*, and its value was explored using the parameter sweep. *Chañarales* and old river bed-floodplain vegetation provide a lower amount of forage for livestock, and they cover a low extension of the landscape, so they were assigned *PVeg3=PVeg4*=0.3. *VegRate*, the *PVeg* decreasing rate with time around settlements, reflects gradual vegetation degradation caused by grazing and wood extraction, which represents the decreasing grass, wood, and total vegetation cover observed around settlements (Goirán et al., 2012; Meglioli et al., 2013). *VegRate* causes a decrease of *PVeg* with time around settlements (in up to $5^{th}$ neighbors, i.e., in a 5x5 pixel neighborhood ). This decrease represents a change in  vegetation types at approximately 2 km from the settlement, as observed in vegetation studies of the region. In order to compare final simulated vegetation with current vegetation data, we made a map of final vegetation using the probability values given by vegetation types (*PVeg*) at the final time step of the simulation. Final *PVeg* values were converted to vegetation types according to the rules defined in Table 2: a $5^{th}$ vegetation type results from the degradation (decreasing *PVeg* around settlements) of shrublands into low cover shrublands; woodlands may be converted into high cover shrublands, and these into low cover, degraded shrublands, according to the reduced *PVeg* at the end of the simulation. The new vegetation type represents degraded areas near settlements, where total vegetation, grass, and shrub cover is low, as a result of vegetation removal by animals and humans (Meglioli et al., 2013). We presume that low cover shrublands were not present at the initial simulated time, because they are generated by continuous impacts of livestock and people on vegetation. Although *PVeg* values in old river bed-floodplain vegetation were allowed to decrease during the simulation, the structure was assumed to remain constant, because this vegetation type is characterized by fast growth species, which grow after occasional flooding, and are not used by local population for forestry products, in contrast to *P. flexuosa* woodlands. In fact, old vegetation maps (Prieto, 2000) describe riparian vegetation in the same places as found today, suggesting that livestock has not significantly changed its structure.

d) *Probabilities assigned to water table depth classes*, *Pwatertab*, were chosen considering the effort of manually building a well, and the need of machinery for greater depths. We divided water table depths in 3 classes, and assigned decreasing probabilities with increasing depths, from class 1 (less than 15 m) to class 3 (more than 25 m) (Table 1). The first and third groundwater depth classes were assumed to have the highest and lowest probability, respectively, so *Pwatertab1*=1 and *Pwatertab3*=0.1. The value of the intermediate class, *Pwatertab2*, was assumed to be lower than the first class, because of the greater effort required and risk of collapse in these sandy sediments. The value of *Pwatertab2* was explored with the parameter sweep. We assumed constant water table depths during the simulation in each pixel because of the low extraction rates given by manual pumping, the negligible local recharge rate given by precipitations (Jobbagy et al., 2011), and testimonies of local settlers, who reported constant water table depths during settlements history.

e) *Probabilities associated with old river beds, Ppaleo.* We assigned two probability classes for *Ppaleo*, for pixels located in and outside old river beds (*Ppaleomax* and *Ppaleomin*).  *Ppaleomax*=1 inside old river beds, and *Ppaleomin*, outside old river beds, was explored with the parameter sweep.

## 2.2.4. Model output

In order to compare simulations with observations, the model produces the following outputs: maps of settlements and vegetation, to provide a qualitative comparison, and histograms of the spatial distribution of settlements, which can be compared quantitatively to observations.

1) Simulated settlement map: the output file of simulated settlements is presented as a 750 m pixel map.

2) Vegetation map: we built a vegetation map that represented the degraded vegetation at the end of the simulations using the final partial probabilities given by vegetation type (*PVeg*) across the grid. These values resulted from the gradual modification of *PVeg* around new settlements, during the course of the simulations. We re-assigned a vegetation class to each range of probability values, attempting to represent the types of vegetation that result from degradation associated with settlement activities (grazing, wood, and firewood extraction). The resulting vegetation map depends on the original vegetation (input grids) and the decreasing partial probabilities on each pixel near established settlements, with the criteria shown in Table 2.

3) Histograms: we plotted three pair correlation histograms of simulated settlements, with the variables that most confidently can be measured in the field: settlement-settlement with pair correlation function, g*(r)*, settlement-road, and settlement-river distances (explained in SOM), averaging the results of the *N* simulations. We then calculated the residuals for each histogram between a simulated (average of *N* simulations) and a reference case with the formula:

$$\sum_{i=1}^{n} (H_i' - H_i)^2$$

where $H'_i$ is the value of histogram <*H'*> at bin *i* for the simulated distribution, and $H_i$ is the value of histogram <*H*> at bin *i* for the reference case.

We used this approach for two reasons: to check the stability of the model, and to compare it with the real distribution of settlements. For the first objective, the reference case is the average of *N-1* simulations (as shown in SOM), and for the second objective, the reference case is the histogram of the spatial distribution of real settlements in the study region. We analyzed the stability of the model including all parameters and input files of the study area, running increasing number of simulations, and calculating the residuals averaged at increasing number of simulations. We chose intermediate parameter values for intermediate distance classes, as detailed in Figure 2. These simulations indicate that steady values are reached for settlement-settlement and settlement-road histograms with approximately 20 to 30 simulations. For settlement-river distances, residuals have small fluctuation with approximately 50 simulations (Fig. 2).

To compare simulations and observations with the histograms and residuals, as explained above, a map of real settlement distribution was obtained from Goirán et al. (2012), which was made with existing records from local government offices (Secretaría de Medio Ambiente Mendoza, 2001a and b), and complemented with settlements detected with Google Earth® (Google, Mountain View, CA, USA, http://earth.google.es) images, considering the presence of corrals or housing structures, in addition to a clear area around them, as evidence of active livestock settlements (supporting kml map). The map from Goirán et al (2012) was re-sampled to 750 m pixels, to compare it with model results. Because people do not depend on government assistance to build wells and settlements, there are not complete and updated records of livestock settlements and their condition (whether they are active or abandoned), so we assumed that all settlements detected with the images were active.

**2.2.5- Evaluating the relative importance of environmental factors with a parameter sweep.**

We assigned a maximum probability value to optimum conditions (*Privermax=Proadmax=Pwatertab1=PVeg1=Ppaleomax=Psettlmax*=1). Based on the results of model stability (Fig. 2 and SOM), and the computational requirements to execute multiple simulations, we executed 30 simulations with each combination of parameter values for the following parameters, corresponding to intermediate distance or aptitude classes: *Privermed*, *Privermin*, *Proadmed*, *Ppaleomin*, *PVeg2*, *Pwatertab2* and *Pset*. Probabilities ranged from 0.1 to 0.9, with a step of 0.1, restricting parameter combinations to those that represented decreasing probabilities with increasing distances from rivers, roads, and surface soil:

*Privermed>Privermin*; *Proadmed>Proadmin*=0.1; *Pwatertab2 > Pwatertab3*=0.1.

The parameter combinations described above were ran with 3 values of *Pset*: 0.12; 0.14; and 0.16, resulting in 373248 combinations. This *Pset* range was selected based on preliminary simulations. We created a Ruby script to execute the cases needed for the sweep, combining all the possibilities for the parameters listed above. Using a mini-cluster of three computers (AMD FX-8350 with eight cores running at 4.0GHz), it took 12 days to run the sweep. Since there are a total of 373248 parameter sets, and for each $N$=30 simulations are needed, there are a total of ~11.2 million simulations required. Perfect parallel efficiency using the timing for a single case (1.7 seconds) would give ~9.1 days. The excess execution time means that, as expected, parallel efficiency is not perfect but fairly good given the extra time needed for memory and disk access when running in parallel. Based on this sweep we have been able to select nearly optimal values for those seven variables, reducing the error for the correlation pair for the road and river. This parameter sweep could be improved in the future by porting the code and the sweeping script to GPGPU (Ino et al., 2009; Motokubota et al., 2011).

We calculated the residuals between the simulations and the real case for each pair correlation function (settlement-settlement, settlement-river, and settlement-road) for each set of input parameters. Using the results of this sweep, we restricted simulation results to those that yielded an average number of settlements similar to the real number of settlements, accepting values within approximately 20% of the target value and therefore including simulations providing an average number of settlements between 400 and 550. As an additional condition, we chose only parameter combinations which lead to a standard deviation lower than 20% of the desired number of settlements (standard deviation of 80 settlements).

We used this subset of simulations to explore parameter values that minimized the residuals between histograms of simulations and observations, plotting parameter values versus residuals, and sequentially excluding values that resulted in maximum residuals.

We then ran 100 simulations with parameter values that yielded minimum residuals, and additional sets of 100 simulations sequentially removing each single factor, to analyze the impact of different factors on the histograms and residuals. We chose to run 100 simulations to ensure we were well beyond the stability condition reached after about 30 simulations.

## 3. Results

With the results from the parameter sweep, we analyzed the combination of parameters that minimized residuals, sequentially excluding combinations with maximum residuals. First, we excluded simulations with *Pset*=0.16, because the resulting residuals were higher than those with the *Pset*=0.12 and 0.14 (Fig. 3). Observing the subset of simulations with *Pset*=0.14 and 0.12, the residuals are sensitive to variations of the three parameters related with surface water availability, *Privermed*, *Privermin*, and *Ppaleomin* (Fig. 4). The other parameters, *PVeg2*, *Proadmed*, and *Pwatertab2* did not cause a marked variability on the residuals (Fig. 4). From this analysis, the following parameter values, which yielded lower residuals, were selected: *Privermed*=0.5; *Privermin*=0.3; *Ppaleomin*=0.2. Although minimum residuals were obtained with *Privermin*=0.4, there was not a combination of this value with *Privermed*=0.5 and *Ppaleomin*=0.2 that satisfied the restrictions of mean and standard deviation of number of settlements, so *Privermin*=0.3 was selected. With the selected subset of parameter values (*Privermed*=0.5; *Privermin*=0.3; *Ppaleomin*=0.2; *Privermin*=0.3, *Pset*=0.14) (Fig. 5), *PVeg2*=0.9, *Pwatertab2*=0.7, and *Proadmed*=0.3 resulted in the minimum residuals.

Running 100 simulations with the parameters selected with the sweep, and mentioned above, the exclusion of the single factor that most increased the residuals was the old river bed, *Ppaleomin*, with a six fold increase, being followed by distance from rivers, *PRiverDist*, with a three fold increase (Fig. 6 and 7). Removing the effect of the road, *PRoadDist* had a slight increase in the residuals, while removing groundwater depth (*Pwatertab*) and vegetation (*PVeg*) did not significantly change the residuals.

The regional pattern of simulated settlements with the optimized parameters visually matches the real settlement distribution, with high densities in the lacustrine plains and near rivers, and more sparse, but not uniform, settlements in the aeolian plain, far from rivers (Fig. 8).

The optimized simulations resulted in an average of 167 settlements (44%) established randomly, mainly in areas of the aeolian plain distant (more than 7.5 km) from rivers, and 213 by maximum probability, mainly near rivers. The histograms for settlement-settlement distances near (up to 10 pixels or 7.5 km) and far (more than 10 pixels or 7.5 km) from rivers show different spatial distributions, with a lower aggregation in the later case (Fig. 9). The high aggregation of real settlements in the first histogram bin in Fig. 9b is given by a cluster of settlements in the NE of the region, more than 7.5 km away from the river. Excluding this cluster, most settlements in the aeolian plain show a low aggregation, and random distribution (Fig. 8). The resulting vegetation map, which includes the degradation given by livestock settlements, shows a similar distribution of vegetation classes as in the real vegetation map, although the reduction of vegetation around settlements, and the total remaining area of woodlands are higher in the model than in the real case (Fig. 10).

**4. Discussion**

The hypothesis, that surface and groundwater are the most important determinants of the spatial distribution of livestock settlements, was partially supported by our simulations, which indicated a clear effect of surface water (rivers and old river bed parameters), but a negligible effect of groundwater depth on settlement spatial distribution. The study by Goirán et al. (2012) indicated areas of settlement aggregation, but the relative importance of groundwater, rivers, old river beds and roads could not be distinguished, because they overlap in several areas. The simulations of settlement dynamics presented in this study provide an evaluation of different environmental drivers of settlements. Parameters related to surface water (rivers and old river bed) were the most important drivers to approximate the real settlement spatial distribution. The residuals reached minimum values only when the probability of the third distance class from rivers (*Privermin*) was less than half the probability of the first distance class (Fig. 4). The probabilities associated with the old river bed also needed to be lower outside the old river bed (*Ppaleomin*) in order to reach minimum residuals. The residuals increased six and three times if the old river bed or river probabilities were removed from the optimized simulations, respectively (Fig. 7). Several ecosystem services are better in the old river bed than in the aeolian plain, such as higher surface water availability, better groundwater quality, and easy access, given by smooth slopes. However, it is possible that present day occupation is affected by a memory of past environmental conditions, although present and past effects are difficult to distinguish. Archaeological studies show most of pre-hispanic occupancy near rivers and lakes, both in the Mendoza and Tunuyán rivers (Chiavaza, 2009 and 2014; García Llorca and Cahiza, 2007; Cahiza and Ots, 2005). Pre-hispanic remains (i.e. fish bones) indicate permanent occupancy and surface water in the main old river bed, (García Llorca and Cahiza, 2007; Cahiza and Ots, 2005: Chiavazza and Prieto Olavarría, 2004). Settlements may persist in the same locations as in the past, although present day conditions have changed. Guanacache wetlands, which have been populated since prehispanic times (Abraham and Prieto, 1991), are almost dried at present as a consequence of geologic and water use changes. However, settlement aggregation persists in these areas to the present. Rivers have lower flows at present, because of upstream use in irrigated oases, and the lakes that they sustained are no longer permanent, because of deep channeling and lower flows. Some settlements are clustered in the river plain, at more than 7 km from the present river channel (Fig. 8) (outside the area of river influence in the present, simulated with the model). These settlements have probably been influenced by the past river channel, which has shifted slightly to the East. The old river bed has been dry since 1778 (Prieto, 2000), but a high aggregation remains in this area, probably because there are no better conditions in the rest of the region to sustain the increasing population. Multimodel studies suggest that in vulnerable regions, climate change will significantly add to the problem of water scarcity that is already arising from population growth, causing domestic instability and migration (Schiermeier, 2013). In our study area, decreasing water availability associated with decreasing river flows is not reflected in changes of occupation patterns. However, decreased water availability probably affects productivity, human-ecosystem interactions, and life quality, so it would be important to plan future development as a function of present and future resource availability.

The construction of the paved road along the old river bed for most of its length was also followed by the provision of electricity and other goods. However, the apparent benefits of the road are not strong enough to cause a significant attraction or aggregation of settlements. This suggests that past and present benefits of the old river bed associated with water availability have a higher impact on settlement establishment than technological and transport benefits given by the road. This is supported by observations of new settlements that were established along the recently built aqueduct, after the input maps for these simulations were elaborated. Water availability is clearly more important to settlers than electricity and transportation.

A finding that did not agree with our hypothesis is the insignificant role of groundwater depth on settlement establishment. Most of the region has groundwater depths that may be accessed by manual metal tools (up to 15 m in interdune valleys). Settlers do not have detailed information about water table depths, so probably they do not include them in their decisions to establish a settlement. Once settlers start to excavate a new well, they may continue although the water table is a few meters deeper than expected. If other factors are favorable for livestock activity, people may invest the necessary effort to reach deeper groundwater, in spite of the difficulty.

Vegetation does not appear to influence settlements spatial patterns, contrary to our hypothesis. Vegetation resources from different vegetation types are probably transported to other areas for construction and fuel, during construction and maintenance periods, while water is needed everyday to keep the animals. People may prioritize water over vegetation "on site" availability, and invest the necessary efforts to transport vegetation goods. As for animal food, goats, the main animals raised in the area, consume shrubs and grasses present in both, woodlands and shrublands. *Prosopis* pods, offered only by woodlands, are collected by the owners and stored for winter reserves, so this food resource may also be transported to the settlements from the surrounding vegetation. The simulated vegetation map resulting from the degradation of the vegetation around settlements has a similar spatial distribution of different vegetation classes as the remotely sensed (real case) vegetation. Low cover shrublands in Figure 10 appear in simulations and remotely sensed data in the NW and NE of the grid. Patches of high cover shrublands are immersed in the area of historic woodlands in both, simulations and observation. However, the areas of extreme classes, such as low cover shrublands (degraded areas) and woodlands (well conserved woodlands) are higher in the simulated map than in the real case. Degradation around settlements seems to be overestimated in the simulations, because we simulated degradation as a constant decrease in partial probability with time. Degradation is a complex process affected by many factors such as temporal variability of precipitation, number and type of animals using the land, and management of the range (Ravi et al. 2010). Data about these factors are not available for the region, so increasing model complexity would not improve our understanding of such processes. The lower cover of woodlands in the real than in the simulated map may be given by woodland clear cutting during the end of the 19$^{th}$ and beginning of the 20$^{th}$ centuries, for railroad and vineyard construction, which was not simulated in the model (Rojas et al., 2009). Another source of error on the simulated vegetation is the initial vegetation map, elaborated from the potential distribution of *Prosopis flexuosa* woodlands (Perosa et al., 2014), which is strongly dependent on the environmental factors found on the remaining woodlands. In order to improve simulations of vegetation dynamics, a better knowledge of the temporal changes of vegetation in the region, as well as other factors affecting degradation (i.e., changes of livestock practices, vegetation-precipitation interactions, wood logging) would be needed.

We also hypothesized that new settlements have to be located near a mother settlement, which is simulated in the model as decreasing probabilities with increasing distances from existing settlements. Our results, which show lower residuals for settlement-settlement distances than for settlement-road and settlement-river distances, support this hypothesis. However, the simulations locate settlements at larger distances than in real cases, as observed in the first and second histogram bins (Fig. 6a). This may be due to the overestimation of land degradation by the model, which decreases probabilities around settlements, acting as a "repelling" force.

The model approximatesthe different settlement densities at a regional scale, with higher densities and more aggregation near rivers and old river beds, as observed in the real case, and sparser settlements, with a random distribution, in areas of the aeolian plane without access to surface water (Fig. 8). In these areas of the aeolian plane (7.5 km from rivers), all the simulated settlements were assigned at random, because total probabilities were lower than the threshold value, *Pset*, indicating a low aptitude of the area (Fig. 9b). The need to assign settlements at random may indicate pressures to settle in unfavorable places, such as socio-political drivers that displaced Huarpe individuals to inaccessible sites in the past, seeking refuge from colonial authorities who relocated them to urban and agricultural areas, or Chilean mines as laborers (Escolar, 2007).

Historically, Huarpe populations inhabited the whole territory of North Mendoza and South San Juan, but the aeolian plain was not permanently occupied during pre-hispanic times (Chiavazza and Prieto Olavarría, 2004). The most productive lands of these provinces were transformed into irrigated oases, and developed for industry and agriculture during the 19$^{th}$ century and to the present. Huarpe individuals and communities may have remained and established in NE Mendoza because the low productivity and difficult access to the lands prevented the expansion of irrigation, agriculture, and urbanization. Social and political factors have been proposed to affect settlements patterns of occupancy, such as family ties (Torres, 2008) and attempts by colonial government to aggregate aboriginals in towns in the *lacustrine plains* (Prieto, 2000). We did not attempt to simulate settlement family relationships, but the relative importance of social, political, and environmental factors proposed to influence settlement dynamics in this area could be analyzed by comparing our results with those from agent based models (Macal and North, 2010; Kohler et al., 2012). Currently, Huarpe descendants inhabit the area and national laws grant them land property rights (Gobierno de la Provincia de Mendoza, 2001; Gobierno de la República Argentina, 1989). Government efforts also tend to support local inhabitants with the construction of an aqueduct and roads to improve access and communication to the area, financial support for tourism projects, and feed subsidies for livestock owners during drought periods (Municipalidad de Lavalle, personal communication). These political changes and a growing population may increase settlement densities and change their spatial distribution, increasing land use pressure, with unknown effects on natural resources. In addition, possible climate changes and increasing water demand in upstream irrigated oases may decrease even further river flows to the region, jeopardizing efforts to reach a sustainable development of the area.

Our model suggests that past and present surface water availability is the main driver of the livestock sector in the region, overruling other apparently important factors, such as roads, with the consequent electricity and communication, vegetation, and groundwater depths. Our model could be used to evaluate the effect of changes on environmental factors, such as water provision by aqueducts, road constructions, and deforestation, on settlement dispersion and vegetation degradation in this and other harsh environments, where woodland conservation efforts, water scarcity, and human activities overlap.

**5. Future directions**

In order to use this model as a management tool for arid areas, it would be valuable to apply and test it in similar areas, such as the dunes in the north of our study region (San Juan province, Argentina), the Kalahari or Arabian deserts, which are also used by pastoralists. The model can also be tested by observing land use changes in the region, such as settlement establishments associated with a recently built aqueduct. Several new settlements not included in the simulations have been established in a well-conserved woodland near the paved road and near a recently built aqueduct. It is clear that water availability encourages settlements, as it has occurred since pre-hispanic times, so the provision of fresh water would likely change settlement distributions and densities. The effect of new infrastructure or land use decisions, such as road construction, deforestation, and water provision could be analyzed with the model, by adding these new features in a grid, and observing the resulting settlement distribution and vegetation degradation. Our study magnified the effect of settlements on vegetation degradation, although it matched the general spatial pattern of vegetation change. Future studies should also include the consequences of different management practices aimed at achieving sustainable use of the environment. This could be achieved with more detailed vegetation monitoring and ecosystem studies, to constrain the vegetation simulation in the model.

**6. Conclusions**

The model simulated regional patterns of settlement distribution as observed on the land currently, probably because the simple model of surface water limitation, in addition to stochastic effects, still holds, in spite of other driver factors. However, 44% of the settlements were established at random, suggesting the existence of other drivers, which could be related to demographic or socio-political pressures. Our simulations suggest that environmental features related to water availability have a strong effect on settlement spatial distribution in our study area. Settlements also affect vegetation, decreasing vegetation cover around livestock settlements. Rivers and the old river bed are the most important factor that explain settlement distribution. Even though the presence of a road, built along the main old river bed for most of its length, provides many additional services to the rural population (i.e., electricity, drinking water, transport to commercial centers), the old river bed seems more important for settlement dynamics. Groundwater depth and vegetation, however, were not found to be important drivers of settlement, contrary to initial expectations, probably because the entire region has accessible groundwater and forage, and certain vegetation products are transported to the settlements from surrounding areas.

Our results imply that changes in water availability and water quality will modify settlement distribution and pressures on the environment. Future water provision, which is crucial for the development of the rural population, should consider the possible effects of increasing settlement densities around water sources and the pressures on the surrounding environment, to ensure long-term sustainability.

**Acknowledgements**

We thank our colleagues from the social and archaeological sciences, Diego Escolar, Pablo Cahiza, María del Rosario Prieto, and Alejandro Tognolli, for invaluable insights about the history and culture of local inhabitants. We also thank Pablo Villagra, Juan Álvarez, and María Laura Gomez for sharing their knowledge about the vegetation and geology of the area. This study was funded with a PFDT grant for S.G., a PICT grant from the ANPCyT (2011-2703), and a PID grant (2010-2014) from SeCTyP (UN. Cuyo).

Fig 1. Study area showing the aeolian plain (without filling), rivers, road, lacustrine systems, irrigated areas, old river beds, and settlements.

Fig. 2. Average residuals vs number of simulations, with $Pset$=0.14, $Ppaleomin$=$Privermed$=$Proadmed$=$PVeg2$=$Pwatertab2$=0.5, $Privermin$=0.1, and other parameter values indicated in Table 1.

Fig. 3. Residuals of the simulations resulted from the parameter sweep with different $Pset$ values. Only parameter combinations that yielded an average number of settlements ($Nsettlements$) between 400 and 550 (similar to the real number of settlements), and a standard deviation ($Nstd$) lower than 80, are plotted.

Fig. 4. Residuals of the parameter sweep with different values of $Privermed$, $Privermin$, $Ppaleomin$, $PVeg2$, $Proadmed$, and $Pwatertab2$. Only combinations with 400<$Nsettlements$>550, $Nstd$<80, $Pset$=0.12, and $Pset$=0.14 are included.

Fig. 5. Residuals of the parameter sweep with different values of $PVeg2$, $Proadmed$, and $Pwatertab2$. Parameter combinations with the restrictions of Figures 3 and 4 (400<$Nsettlements$>550, $Nstd$<80, $Pset$=0.12, and $Pset$=0.14), and optimized parameters values for $Privermed$=0.5, $Privermin$=0.3, and $Ppaleomin$=0.2, are included in this plot.

Fig 6. Pair correlation histograms for distance between settlements (1-settlement-settlement), settlement to river (2-settlement-river), and settlement to road (3-settlement-road ), with the optimized parameters; (a) Simulations with all factors; (b) without old river beds; (c) without river; (d) without road; (e) without vegetation types; (f) without water table simulation. $r$ (pixels) is the distance in pixels, where one pixel = 0.75 km.

Fig 7. Total residuals, averaged after 100 simulations, for the cases presented in Fig. 6. (ALL) Simulations with all factors; (ORB) without old river beds; (RI) without river; (RO) without road;

(Veg) without vegetation types; (WT) without water table simulation.

Fig 8. **a)** Real settlements distribution **b)** Simulated settlement distribution using parameters optimized with the parameter sweep. One representative instance of the $N$=100 simulations was chosen for this figure.

Fig. 9. Pair correlation histograms of settlement-settlement distances for areas near (a) and far (b) from rivers, lacustrine systems, and old river bed, with the optimized parameters. (c) average residuals for both areas.

Fig 10. Vegetation maps for the study area. (a) Initial (input) vegetation one hundred years ago obtained from potential and historic distribution of the vegetation (b) Real present-day vegetation according to a non-supervised classification from Landsat TM (c) Final simulated vegetation including the effect of degradation around simulated settlements, from the same simulation shown in Fig. 8.

**Table 1.** Input parameters and values used in our simulations. Values marked in bold correspond to the optimized values selected with the parameter sweep.

| Parameter | Description, unit | Value |
|---|---|---|
| *dr* | Width of the bin for the g(r) pair correlation function, pixels | 1.0 |
| *Length* | Side of grid, pixels | 150 |
| *Pset* | Threshold of total probability to place a settlement by high probability (in pixel with *Pmax*) | **0.14** |
| *Pthresh* | Threshold of total probability to place a settlement randomly | 0.85 |
| *SettlementMax* | Maximum number of established settlements | 10000 |
| *TimeminSet* | Minimum time between established settlements, time steps. | 1 |
| *TotalSteps* | Number of time steps that the simulations run. | 2000 |
| *AddRoadStep* | Step time when road is included | 1000 |
| *Ppaleomax* | Probability for settlement inside old river beds | 1 |
| *Ppaleomin* | Probability for settlement outside old river beds | **0.2** |
| *Privermax* | Probability for settlement at the first distance bin from a river (*RiverSqrmin* to *RiverSqrmed*) | 1 |
| *Privermed* | Probability for settlement at the second distance bin from a river (*RiverSqrmed* to *RiverSqrmax*)) | **0.5** |
| *Privermin* | Probability for settlement at distances outside *RiverSqrmin* and *RiverSqrmax* | **0.3** |
| *Proadmax* | Probability for settlement at the first distance bin from a road (*RoadSqrmin* to *RoadSqrmed*) | 1.0 |
| *Proadmed* | Probability for settlement at the third distance bin from a road (*RoadSqrmed* to *RoadSqrmax*) | **0.3** |
| *Proadmin* | Probability for settlement at distances outside *RoadSqrmin* and *RoadSqrmax*. | 0.1 |
| *Psettlmax* | Probability for settlement at distances between *SettlSqrmin* and *SettlSqrmax* | 1 |
| *Psetttlmin* | Probability for settlement at distances outside *SettlSqrmin* and *SettlSqrmax* | 0.1 |
| *PVeg1* | Probability for settlement at vegetation class 1 *(Woodlands)* | 1.0 |
| *PVeg2* | Probability for settlement at vegetation class 2 *(High cover shrublands)* | **0.9** |
| *PVeg3* | Probability for settlement at vegetation class 3 (*Chañarales*) | 0.3 |
| *PVeg4* | Probability for settlement at vegetation class 4 (*Floodplain and old river bed vegetation*) | 0.3 |
| *PVeg5* | Probability for settlement at vegetation class 5 (*Low cover shrublands*) | 0.3 |
| *Pwatertab1* | Probability for settlement at the first water table depth class (1 to 15 m) | 1.0 |
| *Pwatertab2* | Probability for settlement at the second water table depth class (16 to 25 m) | **0.7** |
| *Pwatertab3* | Probability for settlement at the third water table depth class (>25 m) | 0.1 |
| *RiverSqrmin* | Minimum distance to a river where bin 1 begins, pixels$^2$ | 0 |
| *RiverSqrmed* | Limit of distance to a river where bin 1 ends and bin 2 begins, pixels$^2$ | 15 |
| *RiverSqrmax* | Limit of distance to a river where bin 2 ends, pixels$^2$ | 50 |
| *RoadSqrmin* | Minimum distance to a road where bin 1 begins, pixels$^2$ | 0 |
| *RoadSqrmed* | Limit of distance to a road where bin 1 ends and bin 2 begins, pixels$^2$ | 15 |
| *RoadSqrmax* | Limit of distance to a road where bin 3 ends, pixels$^2$ | 50 |
| *SettlSqrmin* | Minimum distance to a settlement with the maximum probability for settlement, pixels$^2$ | 0 |
| *SettlSqrmax* | Maximum distance to a settlement with the maximum probability for settlement, pixels$^2$ | 5 |
| *VegRate1* | Rate of decrease of *PVeg* at each time step for first neighbors due to vegetation degradation. | 0.015 |
| *VegRate2* | Rate of decrease of *PVeg* at each time step for second neighbors due to vegetation degradation | 0.015 |
| *VegRate3* | Rate of decrease of *PVeg* at each time step for third neighbors due to vegetation degradation | 0.010 |
| *VegRate4* | Rate of decrease of *PVeg* at each time step for fourth neighbors due to vegetation degradation | 0.010 |
| *VegRate5* | Rate of decrease of *PVeg* at each time step for fifth neighbors due to vegetation degradation | 0.010 |

**Table 2**: Criteria used to elaborate the output vegetation map, at the end of the simulations. The input vegetation maps have only four vegetation classes, with initial *PVeg* values presented in Table 1. The establishment of a new settlement decreases *PVeg* with time in neighbor pixels according to *VegRate*. The final vegetation map is elaborated with the initial vegetation class and the resulting *PVeg* at the end of the simulation, reassigning vegetation classes as described below.

| **Initial vegetation class** | ***PVeg* at the end of the simulation, resulting from vegetation degradation.** | **Vegetation class at the end of the simulation** |
|---|---|---|
| 1 (Woodland) | 0.7 to 1 | 1 (Woodland) |
| 1 (Woodland) | < 0.7 | 2 (High cover shrubland) |
| 2 (High cover shrubland) | 0.7 to 1 | 2 (High cover shrubland) |
| 2 (High cover shrubland) | < 0.7 | 5 (Low cover shrubland) |
| 3 (Chañaral) | all *PVeg* values | 3 (Chañaral) |
| 4 (Old river bed vegetation) | all *PVeg* values | 4 (Old river bed vegetation) |

**Supporting Online Material**
**A- Description of the SeDD model**

    The SeDD model divides the region of interest into a regular grid, made of square pixels, and estimates the "probability" of each pixel of being settled. Probabilities are given by the combination of six factors, assumed to affect settlement establishment and functioning in arid areas: groundwater depth, minimum distance to roads, rivers, and existing settlements, presence of old river beds, and vegetation type. In arid areas, livestock settlements generally rely on surface and groundwater provided by the owners. Low groundwater depths and close distances to rivers may affect the accessibility to owners. Roads and river beds may provide easier access to commercial centers, and different vegetation types may differentially affect settlement establishment and livestock subsistence. These factors are represented in the model in six different layers, with each pixel having a partial probability value for each factor, given by its characteristics and potential services to settlers. For example, a minimum distance to rivers and roads is considered to favor settlements, so maximum partial probability values are assigned to pixels near these features. Certain types of vegetation may provide more services to settlers than others, so they will have higher partial probabilities given by vegetation type. Total probabilities for each pixel are calculated by multiplying partial probabilities given by each of the six environmental factors, assuming that they act independently. The model then assigns a new settlement in pixels with maximum probability if a threshold probability value is surpassed , or at random if the threshold is not reached anywhere. After a settlement is established, the partial probabilities given by "vegetation type" in surrounding $24^{th}$ neighboring pixels gradually decrease, representing vegetation consumption by the livestock of a new settlement. Because of the stochastic nature of the model (i.e., a proportion of settlements established at random), multiple simulations are required for a given set of parameters or initial conditions, and simulation results are then averaged. The cases presented here deal with a square grid, but irregular regions, including non-convex topologies, can be easily accounted for by inscribing the desired area into a larger square grid, and taking undesired pixels off by assigning them zero probability for being settled. These "boundary"

conditions are given by a "mask" grid including information on which pixels have zero probability. Because this is still a regular grid, pixels can be simply indexed, or identified by their (i,j) coordinates within the grid, where "i" represents the row number and "j" the corresponding column.

A flowchart showing how the simulations work is presented in SOM-Fig.1, and explained below:

a) The model first reads the input files and necessary parameters, including rivers, roads, etc. Partial probabilities are evaluated in each pixel for each environmental factor, based on initial values of input grids and parameters. After each time step, these values are updated, including new settlements, and partial probabilities are re-calculated.

b) The probability associated with vegetation type (*PVeg*) is decreased around existing settlements in each time step.

c) Then, the model calculates the total probabilities (*P*) for each pixel in the grid, by multiplying partial probabilities in each pixel. All pairs with maximum *P*, *P= Pmax*, are saved. If there is more than one pixel with the same maximum *P*, the model randomly chooses one of them. Partial probabilities for vegetation type and distance from existing settlements are recalculated at each time step, considering newly established settlements.

d) If *Pmax > Pset* (threshold of total probability *P* to place a settlement by high probability), the model assigns a new settlement in the pixel with maximum *P*.

e) If *Pmax<Pset*, the model may still place a random settlement. First a random number is used to select a single pixel from the grid and, then, a second random number from a uniform distribution in (0,1) is used to decide if a settlement will be established in that pixel. This decision is based on a pre-set rejection threshold (*Pthresh*). For instance, for a rejection rate of 70%, there will be a settlement in the randomly chosen pixel only if the second generated random number is greater than *Pthresh*=0.7. The rejection threshold is related to pressures which force settlements in places where environmental conditions are highly unfavorable.

f) Steps (a)-(e) are repeated for each time step until the end of the simulated time is reached, i.e. by running a number *TotalSteps* of iterations. Then, the model calculates pair correlation functions for the distance between settlements and rivers, settlements and road, and between existing settlements.

g) The user may transform the resulting grid of final *PVeg* values into a final vegetation map, assigning vegetation type classes to given ranges of *PVeg* values.

## 1.1. Required input
**Input Grids**

The landscape information needed by the model to calculate partial probabilities is read from seven input grid files: initial distribution of settlements, existing roads, existing rivers, initial vegetation (five types of vegetation are currently included, which are given different *PVeg* values), water table depth (classified into 3 depth classes), old river beds, and a mask (pixels with total probability=0), which defines the limits of the study area, and can be irregularly-shaped. The mask area corresponds to other productive systems, such as irrigated oases, where livestock settlements cannot be placed, or defined geographic features, such as mountains or rivers, which reduce or prevent animal movements and significant interactions between settlements located in the two sides of these barriers.

**Input parameters and derived variables**

The following list and Table 1 describe each of the model input parameters and derived variables,

which define the size of the grid, the number of time steps, the maximum number of settlements that could be placed, and the probabilities of each class for a given factor.

a) Pixel size, *h (in km)*. *h* does not enter directly in the dynamics of settlements, but has to be taken into account to calculate input and output variables, such as distances from settlements to rivers, roads, and other settlements.

b) *Length (in pixels). N*umber of pixels in a side of the simulated square grid, defines the grid size in pixels**.**

c) Minimum probability value to establish a settlement in pixels with maximum probability (*Pmax*), *Pset*. It determines the minimum environmental requirements to establish a settlement by maximum probability.

d) The rejection threshold, *Pthresh*, determines whether a randomly selected pixel (chosen when *P<Pset*) is settled.

The temporal dynamics of the simulation is controlled by the following *time related parameters.*

e) Total simulated time, in steps, *TotalSteps*.

f) Minimum time elapsed between two settlements, in steps, *TimeMinSet*. Settlements are established only after an elapsed time defined by *TimeMinSet*, in steps. Time steps have to be consistent with vegetation rate changes and with elapsed time between new settlements.

g) Maximum number of settlements that could be established in a given time interval (in this case the total simulated time), *SettlementMax*.

*Partial probabilities* (derived variables) given by spatial patterns, can take values from 0 to 1, and are calculated from the input grids using the spatially related parameters listed below. Note that distance is expressed in units of pixels.

h) Parameters related to the minimum distance between each settlements and a road.

*RoadDist:* closest distance from each pixel to the road.

*RoadDistSqr = RoadDist*RoadDist*

*PRoadDist:* partial probabilities given by *RoadDist.*

*RoadSqrmin, RoadSqrmed and RoadSqrmax* indicate three different distance classes to roads, set in square pixels, that can accept different *PRoadDist.*

The parameters *Proadmin, Proadmed and Proadmax* are probabilities values from 0.0 to 1.0, for each distance class.

*PRoadDist* is then calculated using the following conditions:
   -*PRoadDist =Proadmax if RoadSqrmin < RoadDistSqr < RoadSqrmed*
   -*PRoadDist = Proadmed if RoadSqrmed <RoadDistSqr < RoadSqrmax*
   -*PRoadDist = Proadmin if  RoadDistSqr> RoadSqrmax*

i) Parameters related to the minimum distance between settlements and a river.

*RiverDist:* closest distance from each pixel to a river.

*RiverDistSqr=RiverDist*RiverDist.*

*PriverDist:* partial probabilities given by *RiverDist*.

*RiverSqrmin, RiverSqrmed, and RiverSqrmax* are set in square pixels, and represent three categories of distance to rivers, which can take different *PRiverDist. Privermin, Privermed and Privermax* are probability values from 0.0 to 1.0.

*PRiverDist* is then calculated according to the following rules:
 -*PRiverDist= Privermax if RiverSqrmin < RiverDistSqr < RiverSqrmed*
 -*PRiverDist= Privermed if RiverSqrmed < RiverDistSqr < RiverSqrmax*

-*PRiverDist*= *Privermin* if *RiverDistSqr*>*RiverSqrmax*

j) Parameters related to the distance between settlements.

*SettDist:* closest distance from each pixel to any settlement.

*SettDistSqr*= *SettDist* * *SettDist*.

*PSettDist*: partial probability related to distance between the chosen pixel and the closest settlement, it controls clustering of settlements around "mother" settlements in the following way:

 -*PSettDist*= *Psettlmax* if *SettlSqrmin* < *SettDistSqr* < *SettlSqrmax*

 -*PSettDist*= *Psettlmin* elsewhere.

*SettlSqrmin* and *SettlSqrmax* (in square pixels) determine the range where there is a high probability (*Psettlmax*) of having a new settlement. Outside this range the probability is low (*Psettlmin*). Note that one can choose to have an exclusion area for new settlements around existing settlements by choosing *SettlSqrmin*>0.

*Partial probabilities* given by other environmental features are calculated with the following conditions:

k) Probability given by vegetation type, *PVeg*. Different vegetation types from the input grid are assigned different probability values. In our case, we use five types of vegetation, but more types could be easily added by modifying the code.

l) Rate of vegetation reduction around livestock settlements, *VegRate*. The parameters *VegRate1* to *VegRate5* decrease the *PVeg* during each time step, from nearest neighbor pixels to 5th nearest neighbor pixels from a given settlement, simulating vegetation degradation caused by livestock. There are 5 values for *VegRate* because we take into account a degradation gradient. The pixels closest to the settlement might be degraded differently that pixels further away, so typically *VegRate1*>*VegRate2*>…>*VegRate5*. For a given pixel, *PVeg* (t+Dt)=*PVeg*(t) – *VegRate*, where *VegRate*=0 if there are no close by settlements, and *VegRate*=*VegRate*$_i$, if there is one settlement within a neighbor shell *i*. The effect of several settlements is assumed to be additive, which is a strong but reasonable assumption. This means that *PVeg* for a given pixel will decrease as if the vegetation in the pixel was affected by each neighbor settlement independently of each other.

m) Probabilities given by groundwater depth, *Pwatertab*. Groundwater is defined by three depth classes, with an associated probability for each one. Three classes are enough to describe a typical situation in drylands, where it would be difficult to reach large water-well depths.

n) Presence of an old river bed, *Ppaleo*. Pixels inside and outside the area defined as the old river bed are assigned *Ppaleomax* and *Ppaleomin*, respectively.

Note that one cannot set minimum probability values to zero, because otherwise that pixel would never be settled. One has to choose instead a small but finite values for probabilities like *Ppaleomin* or *Psettlmin,* and in this study we choose to set them as 0.1, while *Ppaleomax* or *Psettlmax* are set to 1.

## 1.2. Simulation Output

The simulation returns several output files, detailed in the list below.

1) Grid of simulated settlements, indicating the position within the grid of old and newly established settlements. The grid indicates by different characters whether a new settlement was established randomly, in an unfavorable pixel, or by maximum probability.

2) Grid of *PVeg* values. This grid presents final probability values (0 to 1) given by vegetation, resulting from the degradation of vegetation around settlements, at a rate given by *VegRate*. This grid can be transformed into a vegetation map, assigning vegetation types to different ranges of *PVeg* values. Decreasing *PVeg* values represent decreasing vegetation cover and grass abundance around settlements, as commonly observed in arid ecosystems (Goirán et al. 2012).

   3) Pair correlation histograms. As a validation tool, the model generates pair correlation histograms that describe settlement spatial distribution (pair correlation functions, g(r), for each distance class, as explained below). In order to compare all histograms from the same or different simulations they were normalized to have unit area, transforming each bin value to a proportion of the total histogram area. There are several histograms built by default at the end of each execution:
a) settlement-road distances, indicating the number of settlements at different distances from the paved road.
b) settlement-river distances, indicating the number of settlements at different distances from rivers.
For each settlement, we calculate the closest distance to a road/river, and add that distance to the corresponding bin in a histogram. This approach allows us to rapidly identify possible tendencies like clustering close to rivers and roads.
c) settlement-settlement distances, showing settlements density according to the distance between them. For these histograms, the model calculates pair-correlation functions g*(r)* as done for atom simulations, where g*(r)* determines thermodynamical properties (Allen and Tildesley, 1987). *g(r)* is also called the radial distribution function (RDF) and is defined as follows: for an average density of atoms, *g(r)* gives the average "local" density at a distance *r* from a given atom. To calculate the settlement-settlement correlation, the model selects one settlement, builds concentric radial bins, i.e. ring-shaped bins of width *dr*, and counts the number of settlements falling in each bin. This is repeated for all settlements, and the histogram is normalized with the area of the corresponding bins, to be consistent with an average density of the system. This approach can rapidly identify exclusion regions, which for atoms correspond to lack of overlap due to their characteristic "size", and can also identify possible short and long range ordering, with short range ordering associated in this case with clustering of settlements likely associated with family ties. We note that this approach is similar to the ecological point pattern analysis of Wiegand *et al*. (1999) and Wiegand and Moloney (2004). A similar approach was used by Winter-Livneh et al. (2010), who applied spatial analysis (Moran's I autocorrelation and Ripley's K-function) and a general linear model to study the ancient Chalcolithic site distribution pattern in the Northern Negev.

**1.3. Calculating average of simulation results and residuals.**
   For a given set of parameters, the output of the simulations changes in different runs due to the stochastic contributions of the model. Therefore, we executed a number *N* of simulations for a given set of parameters, and calculated the average value for each of three output histograms $<H_N>$: settlement-settlement, settlement-road, and settlement-river. The number *n* of radial bins (distance classes) in the histogram, of width *dr,* should be such that $n*dr \approx 1/4$ of the side length of the region studied, to avoid boundary artifacts (Allen and Tildesley, 1987). Then, we calculated the residuals between a simulated (average of *N* simulations) and a reference case (i.e., in this case, the reference case is the average of *N-1* simulations). The residuals of two histograms of correlation functions for settlement-settlement, settlement-river, and settlement-road distances, versus distance with distance bins Δr are calculated with:

$$\sum_{i=1}^{n} (H_i' - H_i)^2$$

where $H'_i$ is the value (fraction of settlements located at distance *i*) of histogram $<H'>$ at bin *i* for the simulated distribution, and $H_i$ is the value of histogram $<H>$ at bin i for the reference distribution. When the residuals change less than 10% we stop increasing *N*. This approach can be used to compare simulated histograms with real data, considering the histograms of the real settlement distribution as a reference case.

**2. Model evaluation using simplified initial conditions**

In order to test the model behavior, we considered several simple initial conditions, where the output should follow simple trends. These conditions separately include each of the different factors that determine partial probabilities. The values for the input parameters are detailed in the corresponding figure (Fig. 1, 2, 3) and in SOM-Table 1. The exact values in the initial input are not crucial here, because any reasonable set of parameters (based on hypothesized behavior) would give qualitatively similar results. In addition, in order to test model stability, 100 simulations were run with the environmental variables and parameter values detailed in SOM-Table 1 for case *i*.

SOM-Fig. 2 shows settlement distributions for different cases (a to i), while SOM-Fig. 3 and 4 show the pair correlation functions of settlement-settlement, and settlement road distances, respectively, for the cases in SOM-Fig. 2.

As an initial test, we made a grid with a road across the diagonal, with one initial settlement close to that road (SOM-Fig. 2a, 3a and 4a).When we considered the distance between settlements (*PSettDist*) as the only driving factor for settlement establishment (all other partial probabilities set to 1), with a maximum distance class of 6 pixels (up to 1.8 km), the settlements are set in a cluster around the first settlement (SOM-Fig. 2b; 3b, and 4b). In SOM-Fig. 2c, 3c, and 4c, we only changed the distance between settlements, *SettlSqrmin* and *SettlSqrmax*, to 8 and 24 pixels respectively (2.1 km and 3.6 km). The resulting settlement cluster is larger and the distance amongst settlements increased (SOM-Fig. 3c).

In the cases presented in SOM-Fig. 2d and 2e, we added the distance from the settlement to the road as a driving factor (*PRoadDist*), and the settlements are set alongside the road from both sides, while the distance between settlements is maintained.

In SOM-Fig. 2e the settlements could not be established near each other, because the minimum distance between them (*SettlSqrmin*) was set to 2.1 km, and the maximum distance between settlements was also modified (*SettlSqrmin*=3.6 km). The settlements are more dispersed along the road compared with SOM-Fig. 3.d, and the distance between them has increased (SOM-Fig. 3e and 4e).

In the next simulation, SOM-Fig. 2f, we used the same input from case d, adding some settlements at random, with *Pset*= 0.5 and *Pthresh*= 0.1.

The simulation in case g (SOM-Fig. 2g) added the same type of vegetation in all the grid and vegetation reduction rates around settlements (*VegRate*), also taking into account the distance between settlements. In case h (SOM-Fig. 2h), the simulation only took into account vegetation types (*PVeg*) and vegetation reduction rate (*VegRate*), and the settlements were not bounded by the distance between them or to the road (SOM-Fig. 3h and 4h). SOM-Fig. 5 shows the degradation of vegetation, expressed in reduced *PVeg*, around settlements, according to *VegRate* values of case g and h.

The simulation in case i (SOM-Fig. 2i) considered the distance between settlements, the distance from the road and a random placement of settlements, as controlling factors. The settlements could be placed next to the road and with a minimum distance between them (*SettlSqrmin*) of 0.75 km and a maximum distance (*SettlSqrmax*) of 3.6 km, distance to the road was also increased to a maximum of 10 km (*RoadSqrmax*).

We executed additional simulations to test the model behavior. We placed a river across the middle of the grid, in vertical and horizontal positions, and both cases gave equivalent histogram output, as expected. In another simulation, when we placed vegetation only in the upper middle of the grid, all settlements were placed in that area, according to the model structure. If we half the values of the variables *Pset* and *Pthresh* compared with the values they had in the previous simulation, the number of settlements approximately doubled in both cases (due to the randomness of the simulations the number of settlements could not be exactly doubled). In a simulation when the settlements were placed at random with no environmental factors involved, the results in the settlements-settlement histogram show that the settlements could be placed anywhere in the grid. This gives just a step function with an exclusion of length *h (pixel size)* before the step since settlement-settlement cannot be smaller than *h*.

All of the above simple simulations indicate that the model works as intended, changing settlement patterns according to different environmental forcings.

The number $N$ of simulations required to get stable results, considering the environmental factors in SOM-Fig. 2i, is approximately 60, as shown in SOM-Fig. 6. However, N could be different for other parameter sets.

**4. Model applications and caveats.**

The simulation code is relative small and simple, and it could be easily expanded. New environmental factors can be added by editing and adding a few new lines to the code. The input grids could be elaborated from topographic, hydrogeological, vegetation and settlement maps, often available from research and government institutions for other regions. For instance, elevation maps can be obtained using SRTM-DEM, freely available from NASA (http://www2.jpl.nasa.gov/srtm).

One of the computationally costly steps in the model is to evaluate the vegetation changes in a pixel. This is equivalent of including "long range" effects on the evolution of grid points, which is an extremely challenging problem. Most grid evolutions are carried out for nearest neighbors or at most for second nearest neighbors, for instance for the Ising model of magnetism (Binder and Heermann, 2010) and the Plant Spread Simulator (Fennell et al., 2012). Here we are including up to $5^{th}$ neighbor shells, which adds up to 24 neighbors, which is equivalent to consider all neighbors within a square of side 5 *h*, for a pixel at the center of that square. If our pixel size is *h*, the $5^{th}$ shell of neighbors is at a distance $((8)^{0.5} h) \sim (2.83 h)$, which for *h*=0.75 km, results in a radius of influence of ~2 km. This is appropriate to model dyrlands used for subsistence livestock production, based on observations of vegetation around settlements (Goirán et al., 2012; Ringrose et al., 1996). To track the evolution of the vegetation with a higher precision and spatial resolution, a higher number of shells would be required. For this purpose, different neighbor tracking methods, which are efficient for large number of neighbors (Allen and Tildesley, 1987), could be implemented.

We note that boundary conditions have to be considered with care. Here we disregard any influence of the sites outside the simulated area, which could be unrealistic in certain cases. One future addition to the model could include a "buffer" boundary region, allowing settlements and conditions outside the simulated area to affect those inside the area.

We did not aim to simulate a "realistic" time evolution, but only settlement distribution at a final time step, because there is no information about the temporal dynamics of settlements to evaluate such aspect of the model. Temporal dynamics could include a time dependent threshold, responding to different environmental pressures. Another way to change the settlement dynamics would be to include an additional random process: when all probabilities have been calculated, add a settlement at a maximum probability site or into another (no so favorable) site, with the decision being taken based on a random number. This would be somewhat similar to a Metropolis Monte Carlo simulation (Allen and Tildesley, 1987) and might provide a more realistic time evolution.

**B-Computational considerations**

The code only requires the GCC compiler and standard compilation tools found in any GNU/Linux distribution. We do not use external libraries that would require additional work to make the simulation run. The simulation is run from the command line. Input parameters have default values placed in the code, which are used if the program does not receive new input parameters. To obtain averages and residuals from multiple simulations, we created a script to execute the simulation for multiple test cases with possibly different input files. This script, which is written in Bash (Ramey and Fox, 2009), runs the simulation $N$ times with any number of input files. For each input file it creates a directory with all the output files of each execution, then saves in a single .csv file each of the pair correlation histograms for each simulation.

The code takes only 1.92 seconds for a typical execution, in a single core of a PC (AMD Phenom x6 1055T 2.8 GHz, 12 GB RAM), for a grid size of 150x150 pixels and during 2000 steps, using only 4 MB of memory. Of this time, 0.34 seconds are used to read input data, 1.51 seconds to evolve the grid, and 0.06 seconds to output results. For a region of 100x100 km, with a grid size of 0.1 km, using matrices of 1000x1000 pixels is still possible to run the code without incurring in memory issues, since it uses only 125 MB of memory. However, the area of vegetation decrease around settlements would be reduced to 400 m, which is unrealistic for the cases which we desire to study, but might be appropriate for other scenarios. The above timing is for a single execution ($N$=1), but about $N$~100 simulations are needed to obtain appropriate statistics for one set of parameters. The current simulation code works using one CPU core, but we also created a script to run N different cases in multiple cores of a single CPU.

A parameter sweep would be needed to test the stability of the results to small changes in the known parameters, or to optimize certain unknown parameters, minimizing the overall error, as explained in the article. A typical parameter sweep might require millions of runs to sample different parameter sets, because a single set will require $N$ repetitions to obtain reasonable statistics, with N~100. Therefore, we implemented a computational script in the Ruby language (www.ruby-lang.org/en/) to speed execution. For instance, if we have to optimize $m$ variables, and each variable can have $z$ different, discrete, values, the number of executions needed for the parameter sweep will be $N\ z^m$. The Ruby script executes $N$ simulations for each set of input parameters and calculates the residuals between each case and a reference case, saving the results in an output file. The script supports multiple CPU cores in a single workstation, queuing one independent process with a given input data in each core, which can provide a large speed-up in code execution in multicore machines.
Improving the speed of the simulation would allow a much broader parameter sweep, and it could minimize errors thanks to improved parameter selection. There are several strategies to improve code performance, but all of them would require significant program additions. For instance, as an alternative to the Ruby script, the code could be modified to run faster using multiple cores in one workstation with OpenMP (OpenMP, 2014). Adding MPI capability, for instance in its OpenMPI flavor (OpenMPI, 2014), to use a cluster of CPUs is also possible. Another approach for improving performance would be to port the code to OpenCL (OpenCL, 2014) or CUDA (CUDA, 2014) to execute the simulation in a GPU (Millán et al., 2010; Preis et al., 2009) speeding up execution running different parameter sets in different GPU threads. (Ino et al., 2009; Motokubota et al., 2011).

**SOM Figures and Table**
**SOM-Figure 1**. Flow diagram of the model.
**SOM-Figure 2.** Simulations with simple initial conditions to test the model. Pixel size is 0.75 km (a) Initial test, with a road (line) across the grid and a single settlement as a black dot. **(b)** distance between settlements *(PSettDist)* as the only driving factor for settlement establishment. *Psettlmax*=1 between *SettlSqrmin*=1 and *SettlSqrmax*=6 pixel$^2$. **(c)** equal to (b) but with *Psettlmax*=1 between *SettlSqrmin*=8 and *SettlSqrmax*=24 pixel$^2$. **(d)** distance from the settlement to the road (*PRoadDist*) also included as a driving factor. *Psettlmax*=1 between *SettlSqrmin*=1 and *SettlSqrmax*=6 pixel$^2$; distance intervals from road correspond to *RoadSqrmin*=0 pixel$^2$, *RoadSqrmed*=15 pixel$^2$ and *RoadSqrmax*=50 pixel$^2$, with *Proadmax*=1.0, *Proadmed*=0.5, and *Proadmin*=0.1, respectively. **(e)** similar to (d) but with *Psettlmax*=1 between *SettlSqrmin*=8 pixel$^2$ and *SettlSqrmax*=24 pixel$^2$. **(f)** *SettlSqrmin*, *SettSqrmax*, and *PRoadDist* as in (d), but adding random settlements, with *Pset*= 0.5 and *Pthresh* =0.1. **(g)** *Psettlmax* =1 between *SettlSqrmin*=0 and *SettlSqrmax* =30 pixel$^2$, adding a constant vegetation in all the grid, and with *VegRate1* and *VegRate2*=0.010, and *VegRate3*, *VegRate4* and *VegRate5*=0.005. **(h)** Settlements are determined only by vegetation decrease (decrease of *PVeg* given by *VegRate*), neglecting all other factors. **(i)** All previous factors (road, settlements, random settlements, vegetation) included, with *SettlSqrmin*=1, *SettlSqrmax*=24 pixel$^2$, and distance intervals from road *RoadSqrmin*=0, *RoadSqrmed*=15 pixel$^2$, *and RoadSqrmax*=178 pixel$^2$.
**SOM-Figure 3.** Pair correlation histograms of settlement-settlement distribution, for the same simulations as in Fig. 2. *r* is distance to settlement in pixels.
**SOM-Figure 4** Pair correlation histograms of settlement-road distribution, for the same simulations shown in Fig. 2. *r* is distance to road in pixels.
**SOM-Figure 5 (a)** Degradation of vegetation for simulation of figure 2.g., which shows the degradation around aggregated settlements **(b)** Degradation of vegetation for simulation of figure 2.h., showing degradation in sparse settlements. Color scale indicates final *PVeg*, resulting from decreasing *PVeg* around settlements according to *VegRate*. Black pixels indicate severe degradation around settlements (*PVeg* 0.3 or lower). A value of 1.0 indicates no degradation.
**SOM-Figure 6.** Average settlement-settlement and settlement-road residuals vs simulation number, *N*, for simulations with the parameters for case *i* (Table 1, Figure 2i).

**Table 1:** Parameter values for the different simulations used to test the model.

| *Cases* | *a* | *b* | *c* | *d* | *e* | *f* | *g* | *h* | *i* |
|---|---|---|---|---|---|---|---|---|---|
| *TotalSteps* | 0 | 200 | | | | | | | |
| *Pset* | 0.5 | | | | | | | | |
| *Pthresh* | | 1 | | | | 0.1 | | 0.8 | |
| *RoadSqrmin* | 0 | | | | | | | | |
| *RoadSqrmed* | | 8 | | | | 15 | | | |

| | | | | | | | | |
|---|---|---|---|---|---|---|---|---|
| *RoadSqrmax* | colspan=7 | 50 | | | | | | 178 |
| *Proadmin* | colspan=2 | 1 | | colspan=3 | 0.1 | | 1 | 0.1 |
| *Proadmed* | colspan=2 | 1 | | colspan=3 | 0.5 | | 1 | 0.5 |
| *Proadmax* | colspan=8 | 1 |
| *Psettlmin* | colspan=8 | 0.1 |
| *Psettlmax* | colspan=8 | 1 |
| *SettlSqrmin* | | 1 | 8 | 1 | 8 | 1 | 0 | 1 |
| *SettlSqrmax* | | | | | | | | |
| *VegRate1* | colspan=5 | 0 | | colspan=3 | 0.01 |
| *VegRate2* | | | | | | | | |
| *VegRate3* | | | | | | | colspan=3 | 0.005 |
| *VegRate4* | | | | | | | | |
| *VegRate5* | | | | | | | | |
| *PVeg1* | colspan=8 | 1 |
| *PVeg2* | | | | | | | | |
| *PVeg3* | | | | | | | | |
| *PVeg4* | | | | | | | | |
| *PVeg5* | | | | | | | | |

**Figure 1**

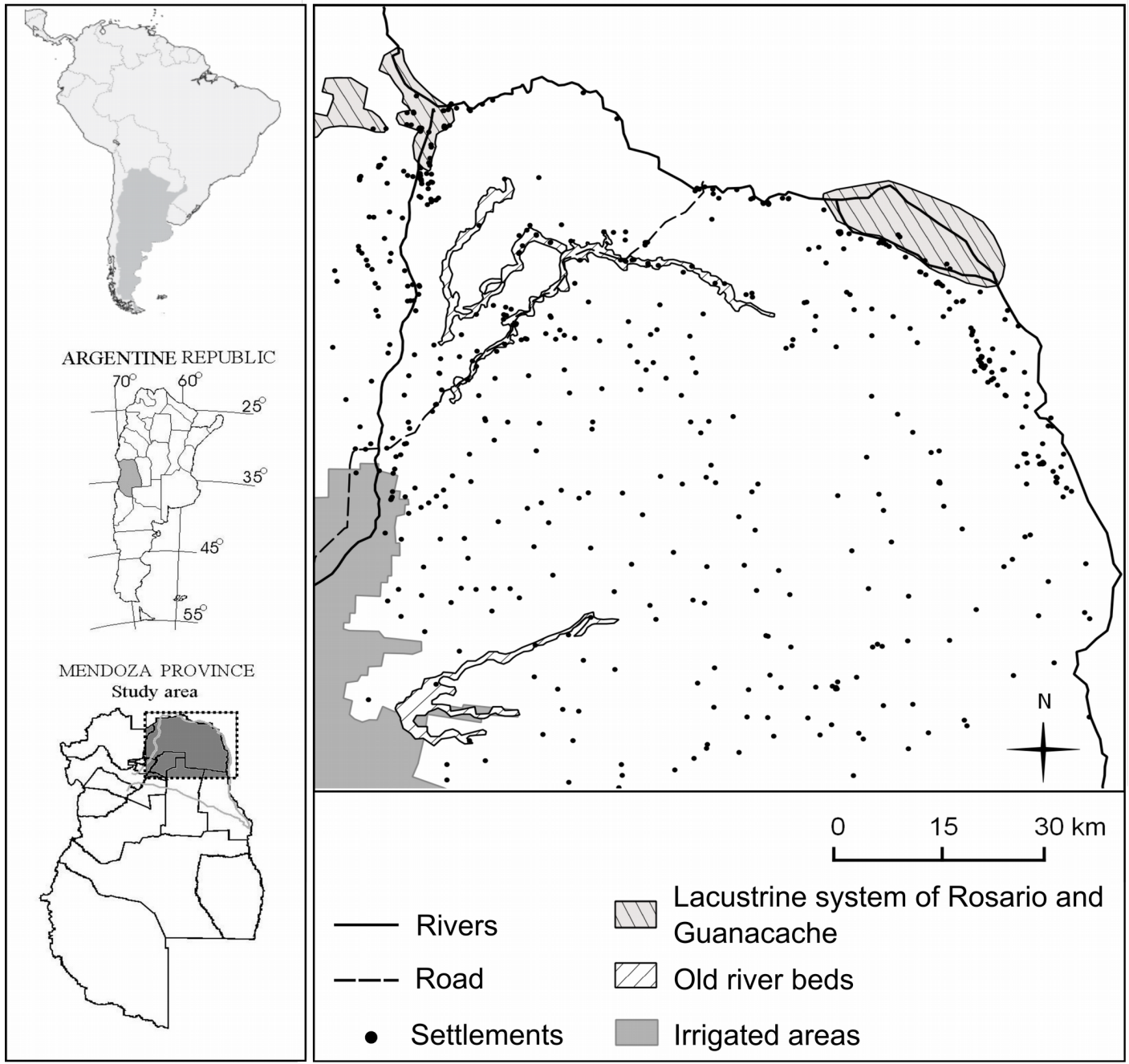

**Figure 2**

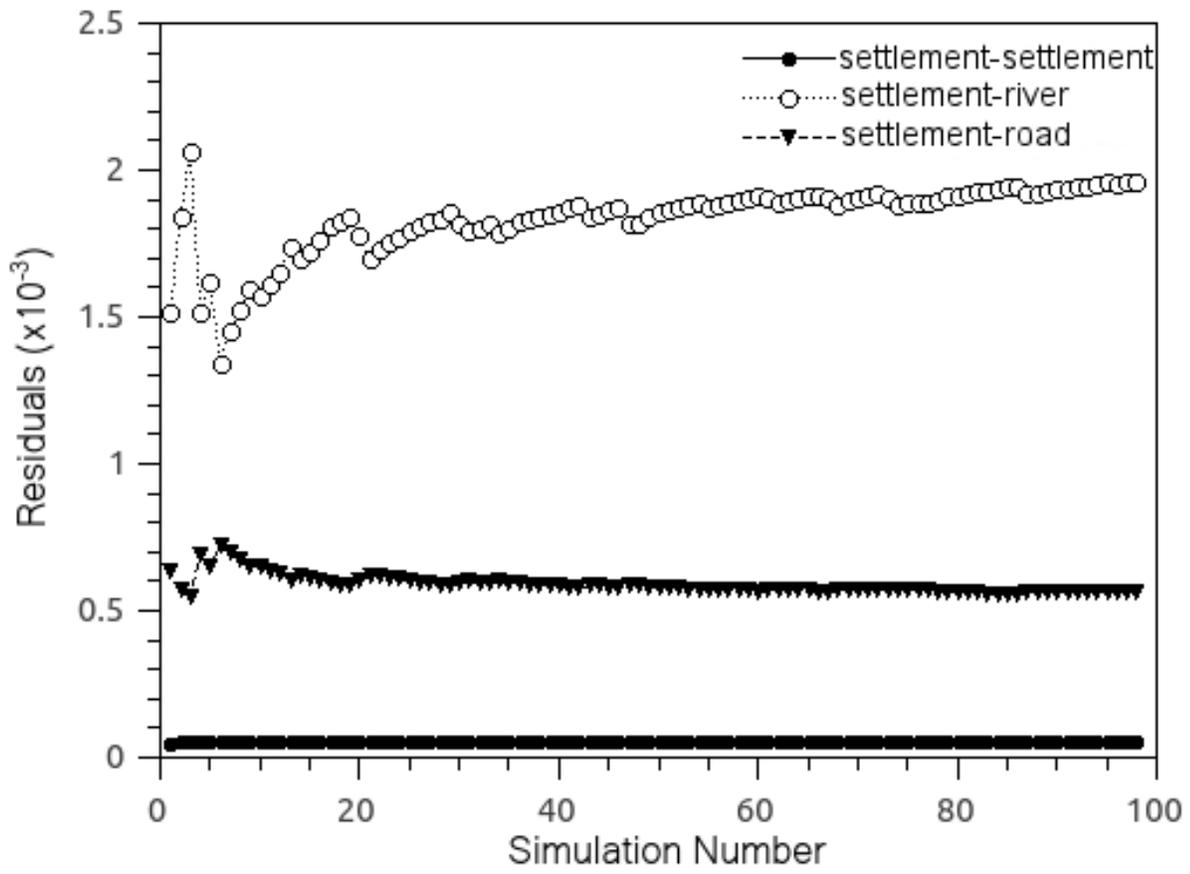

**Figure 3**

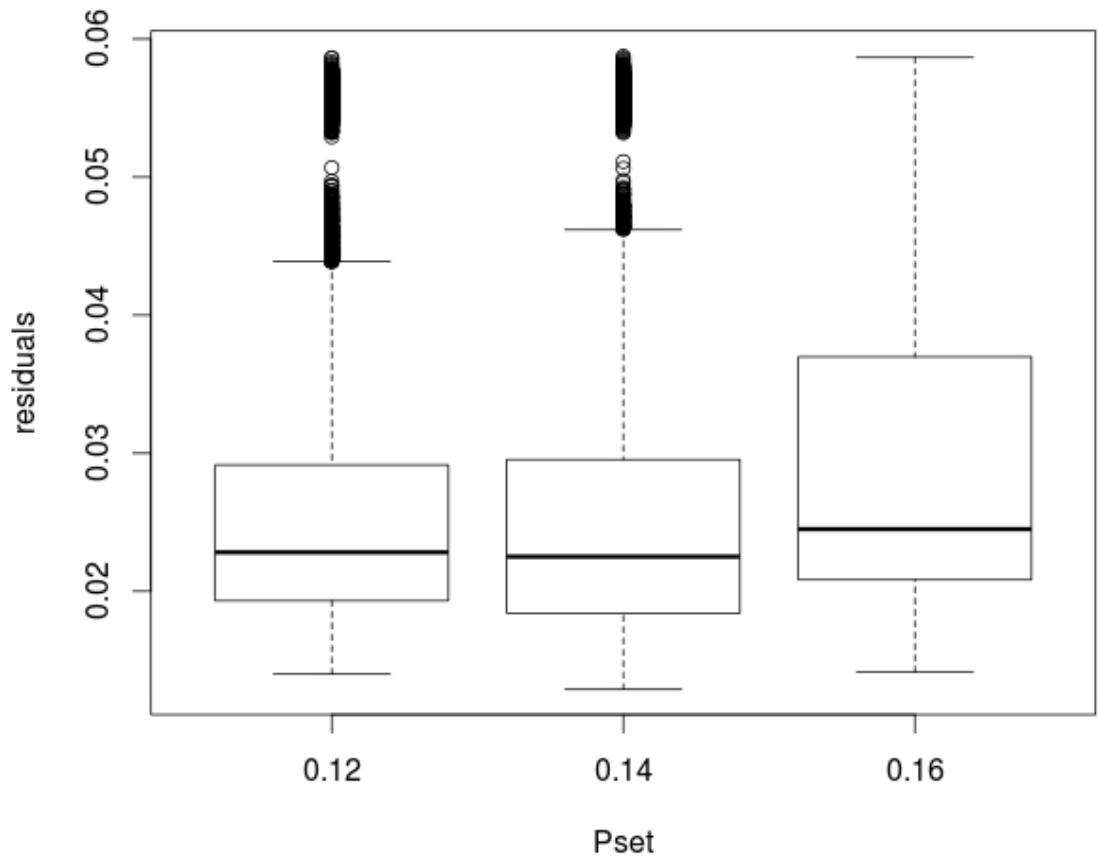

**Figure 4**

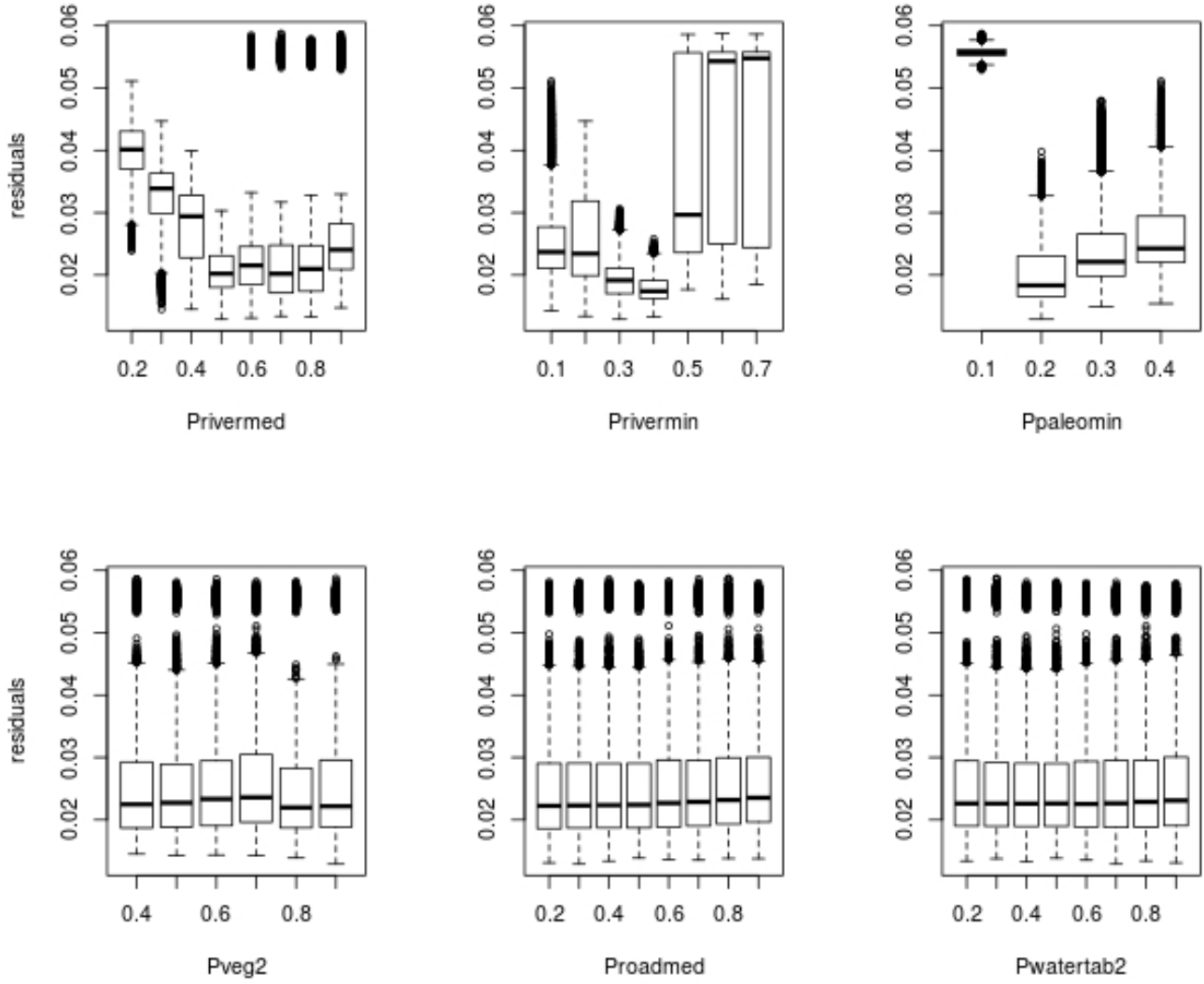

**Figure 5**

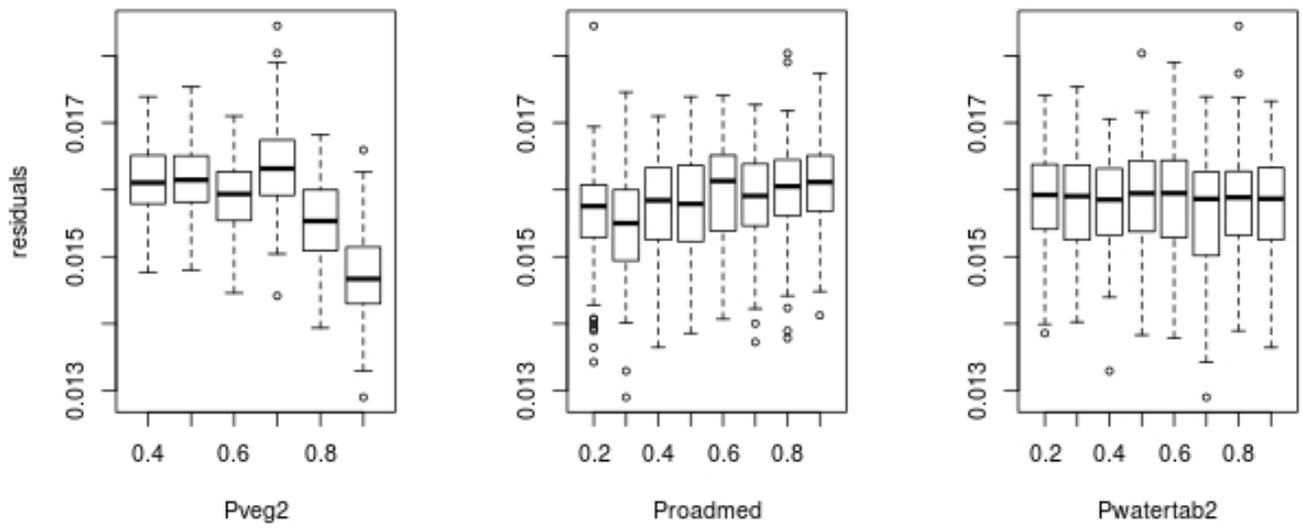

**Figure 6**

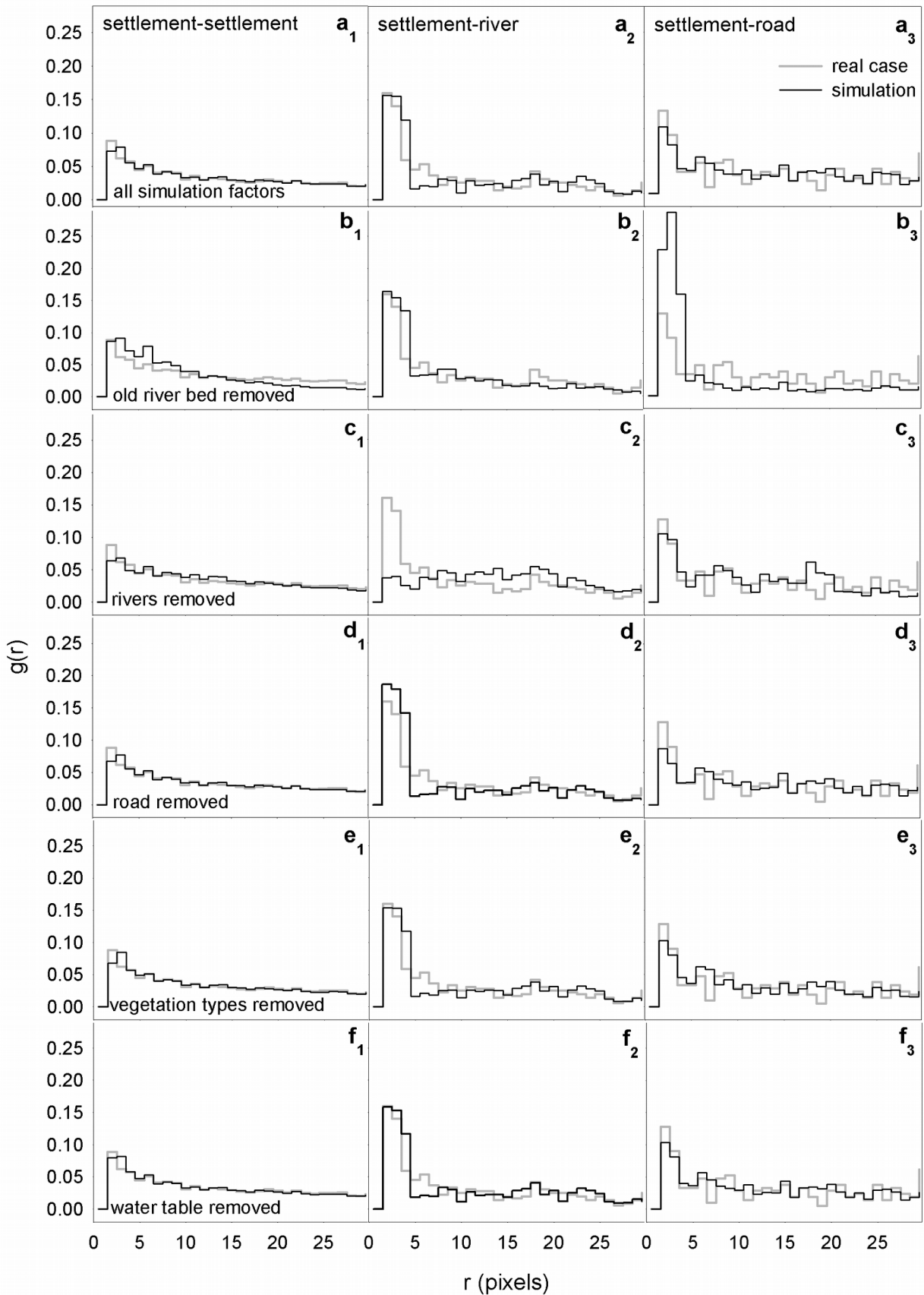

**Figure 7**

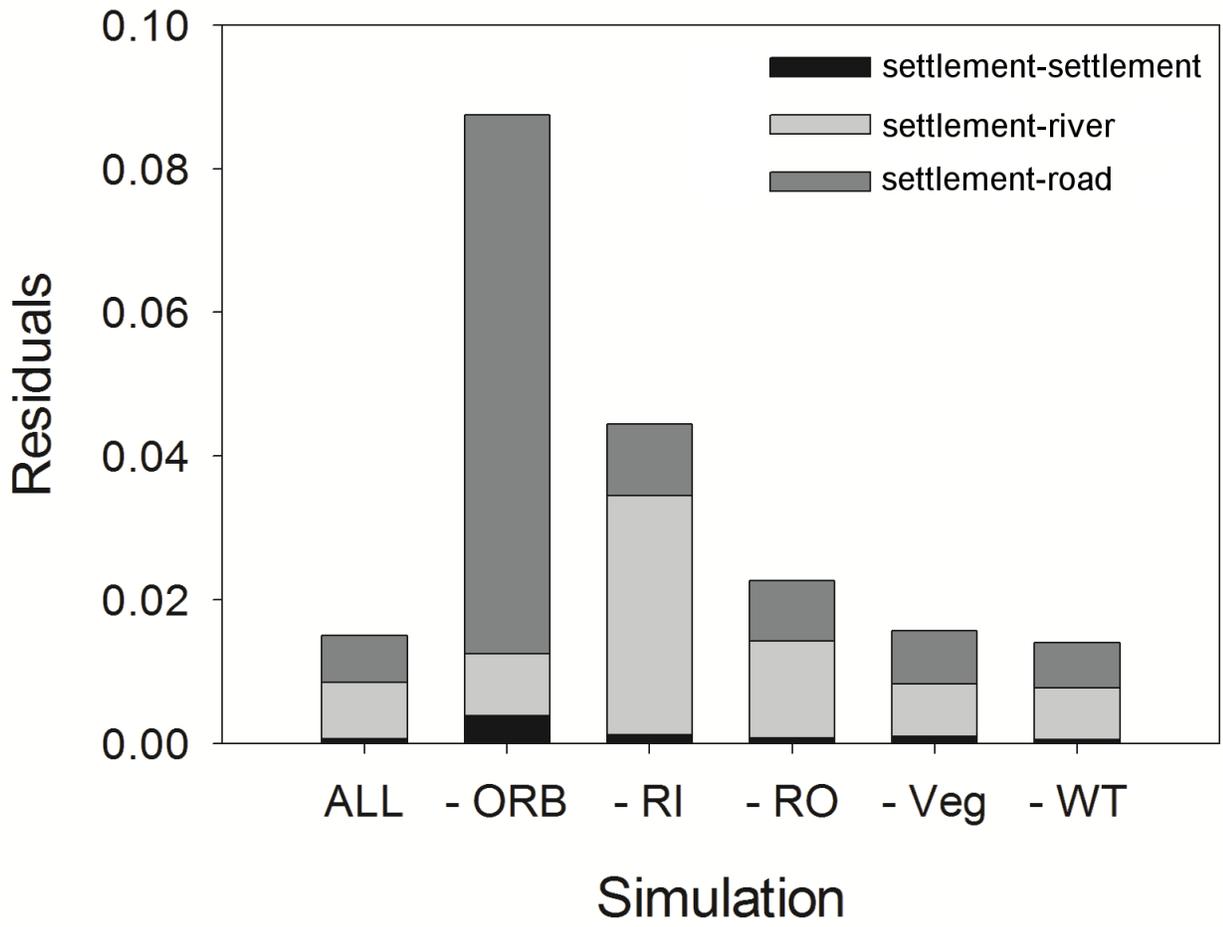

**Figure 8**

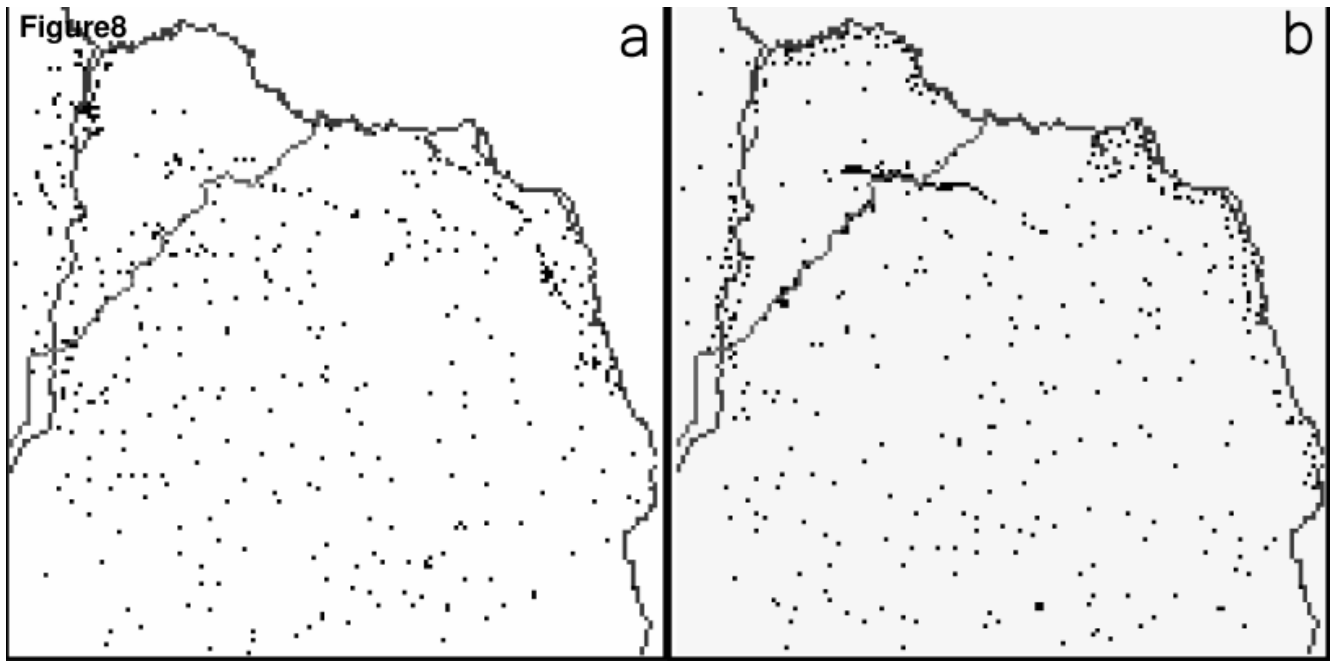

**Figure 9**

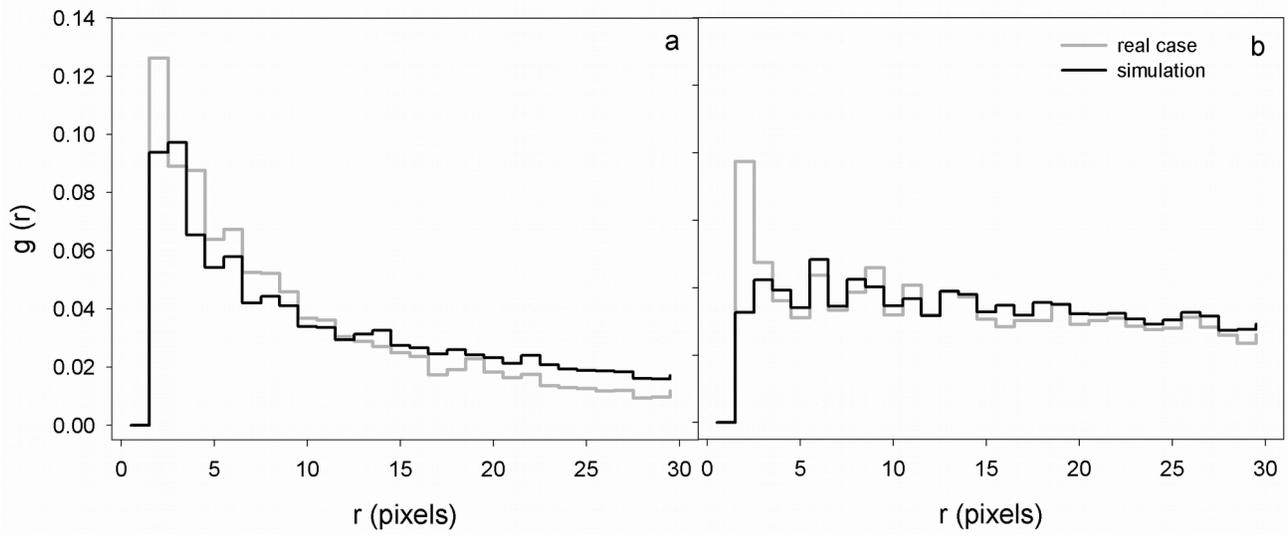

**Figure 10**

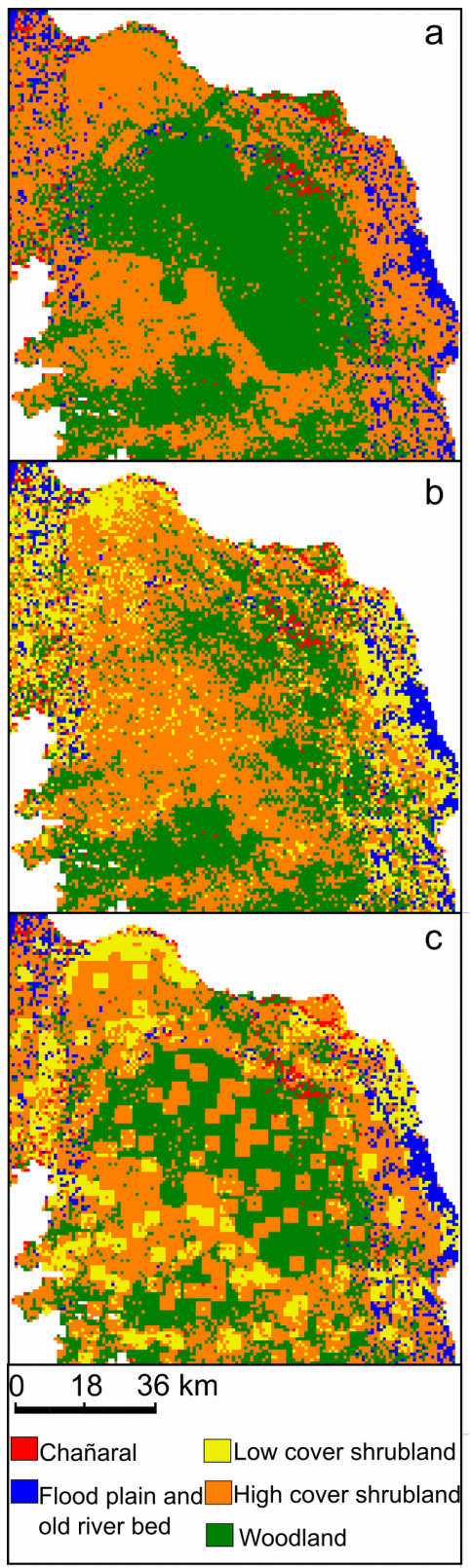

**SOMFig1**

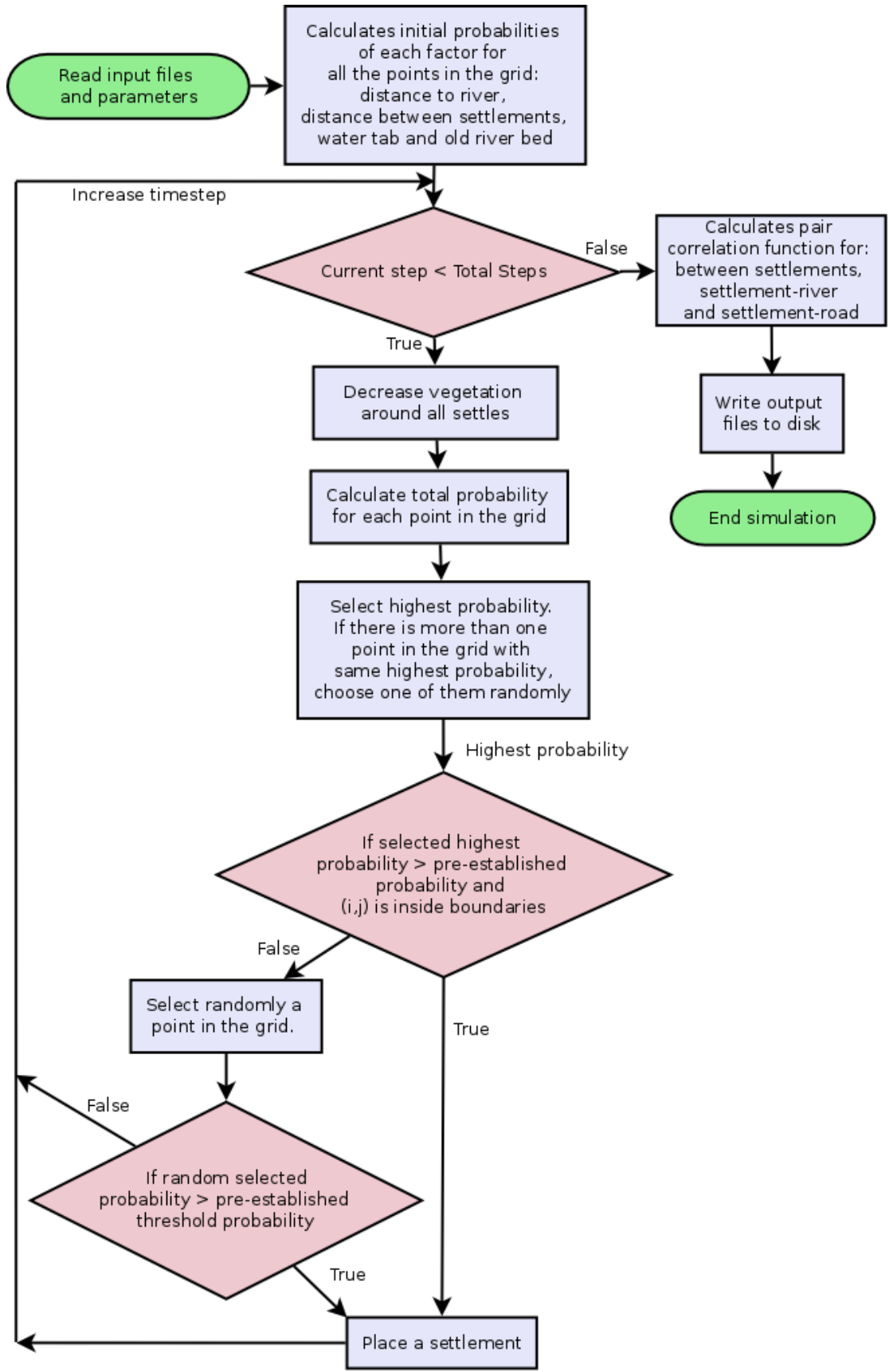

**SOMFig2**

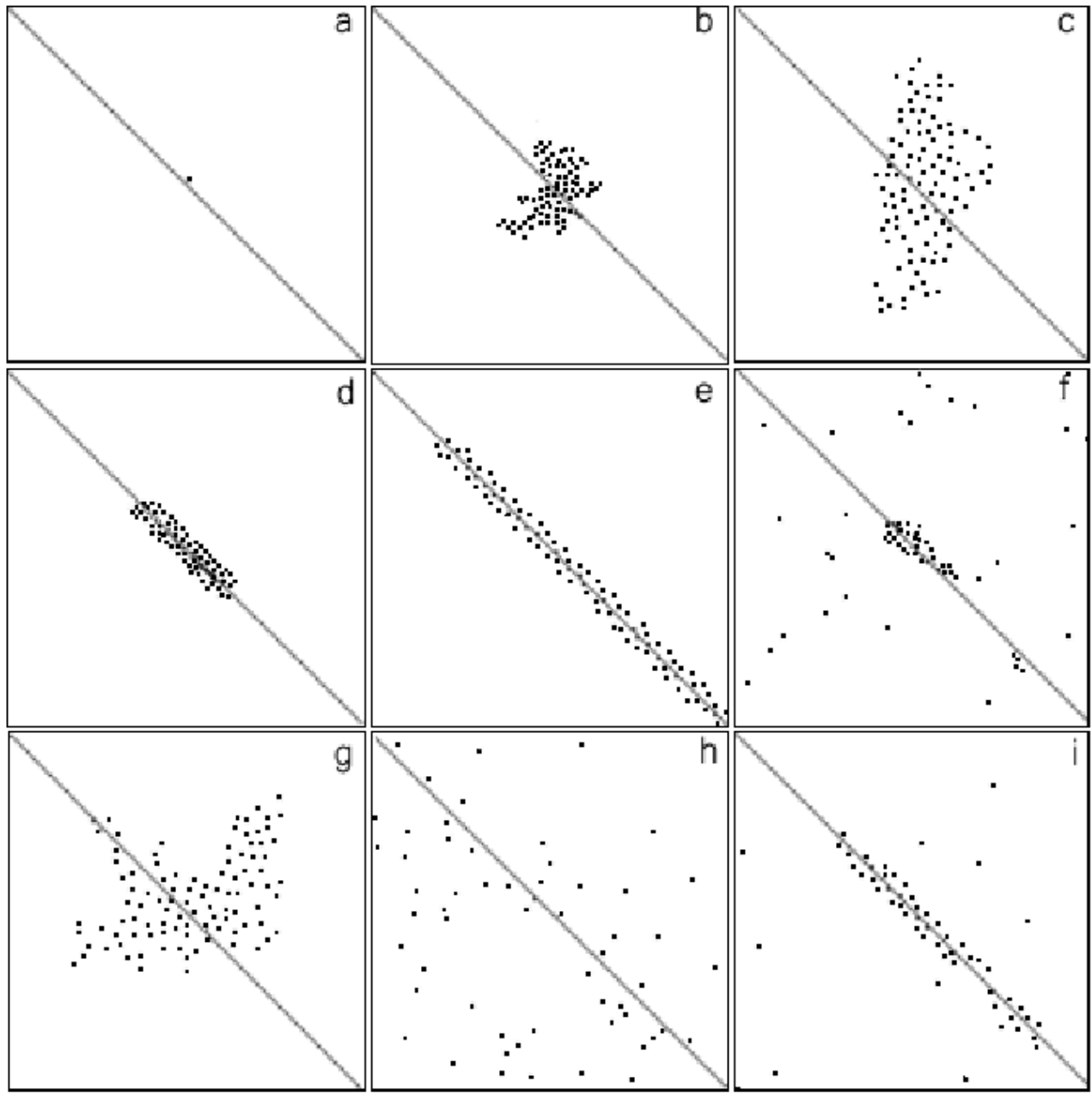

**SOMFig3**

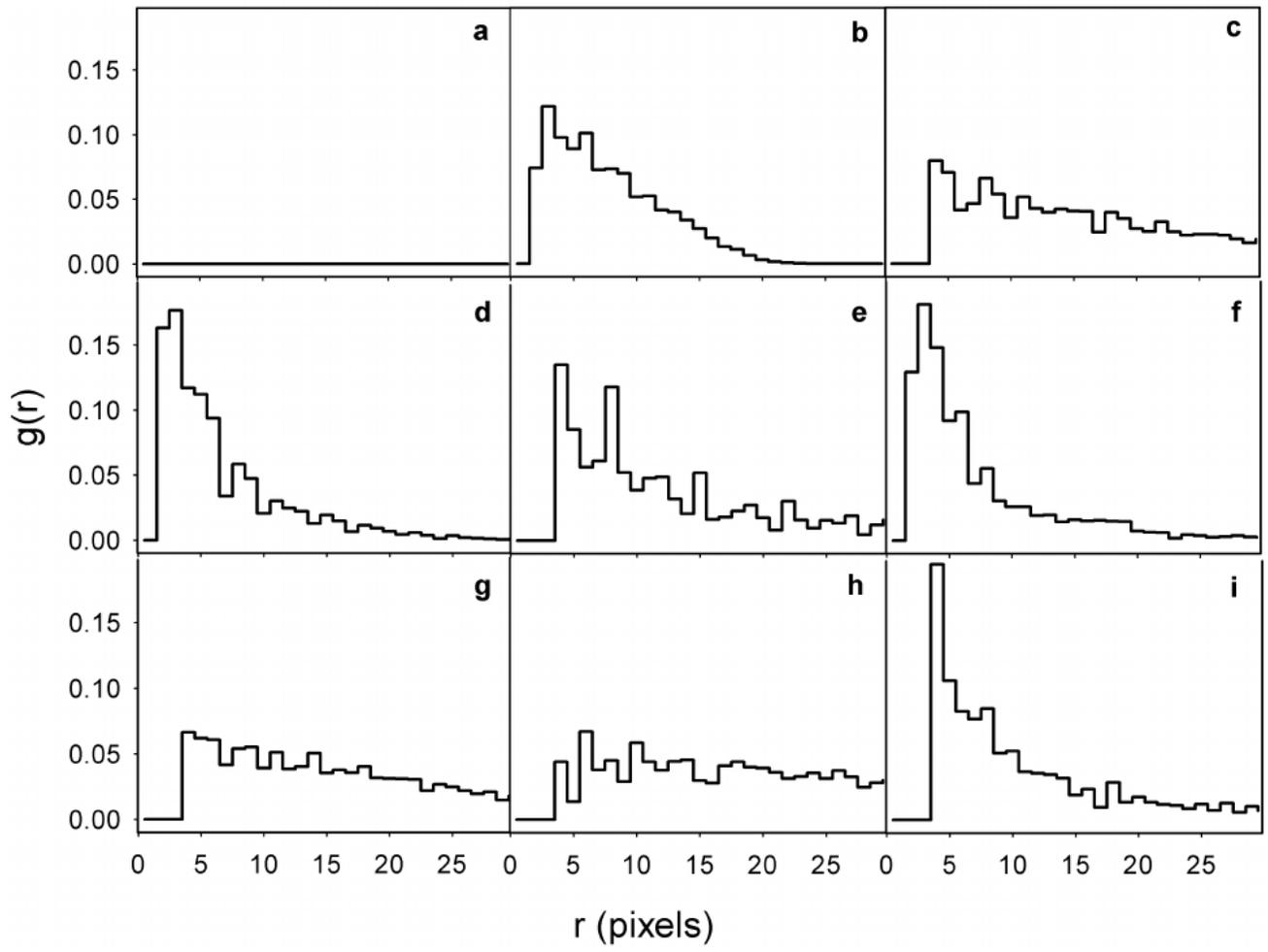

**SOMFig4**

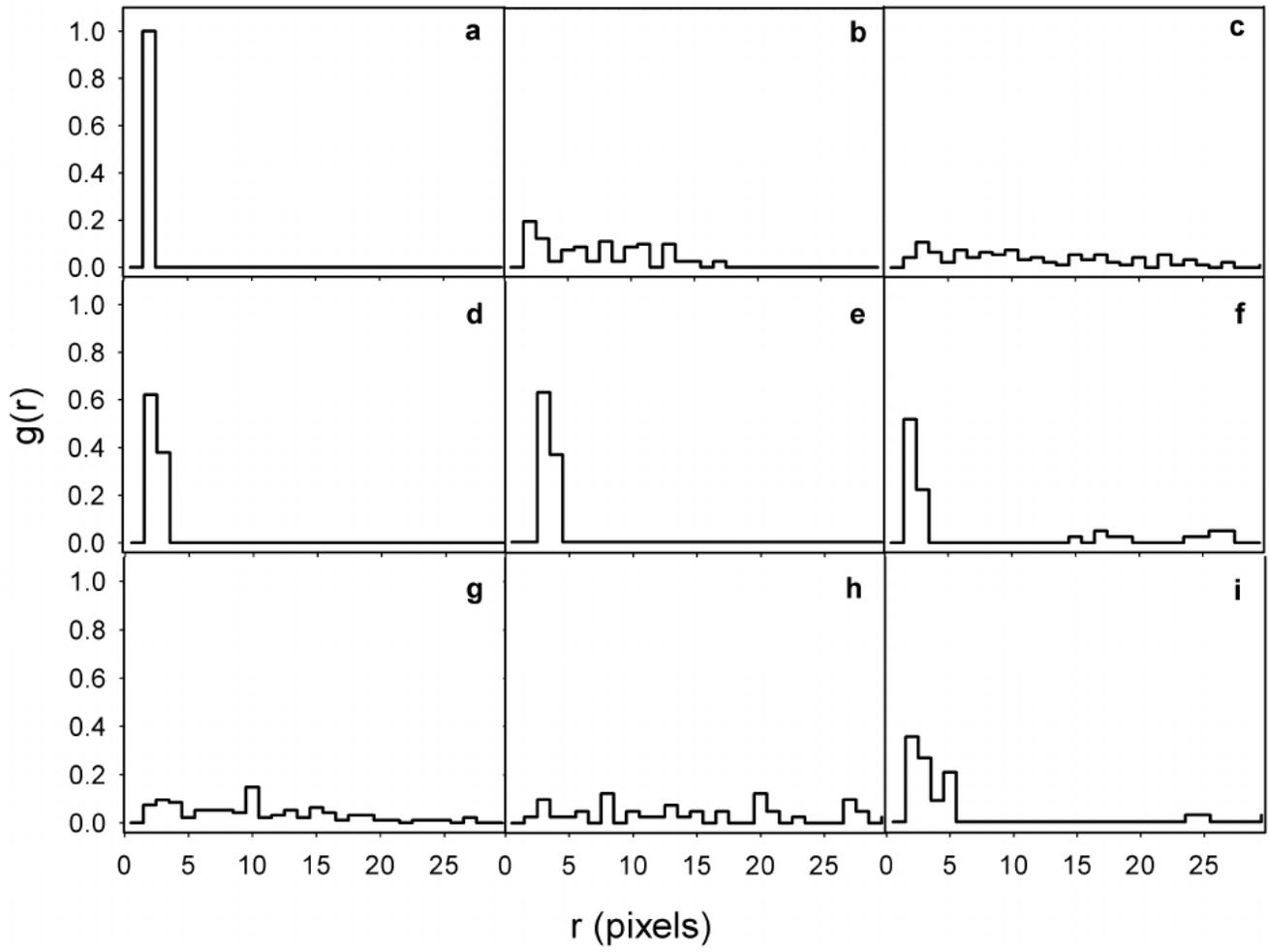

**SOMFig5**

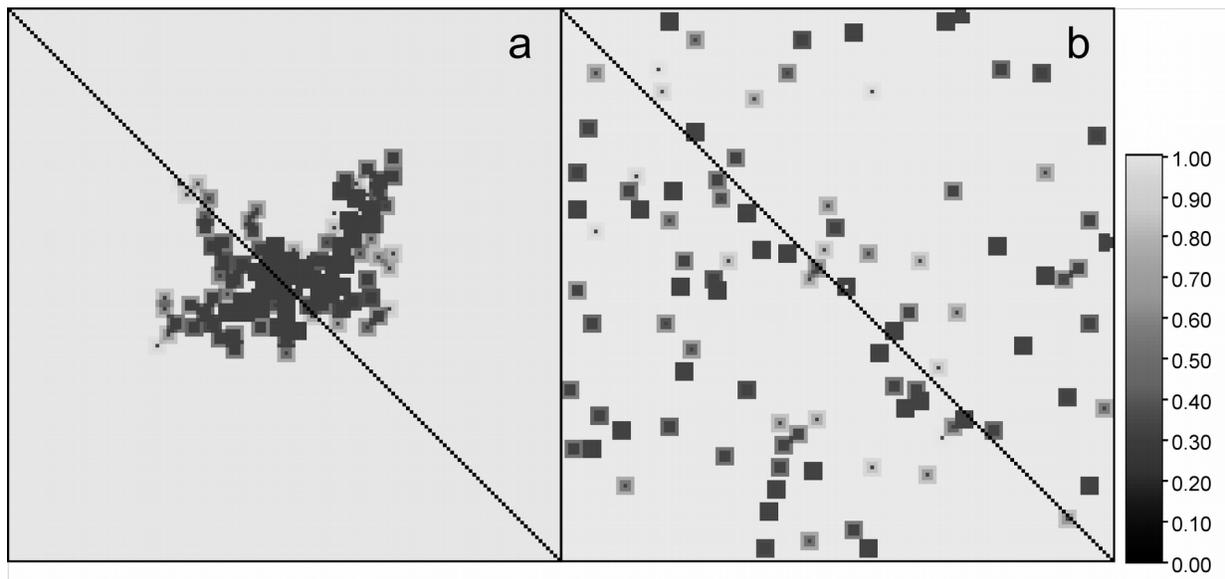

**SOMFig6**

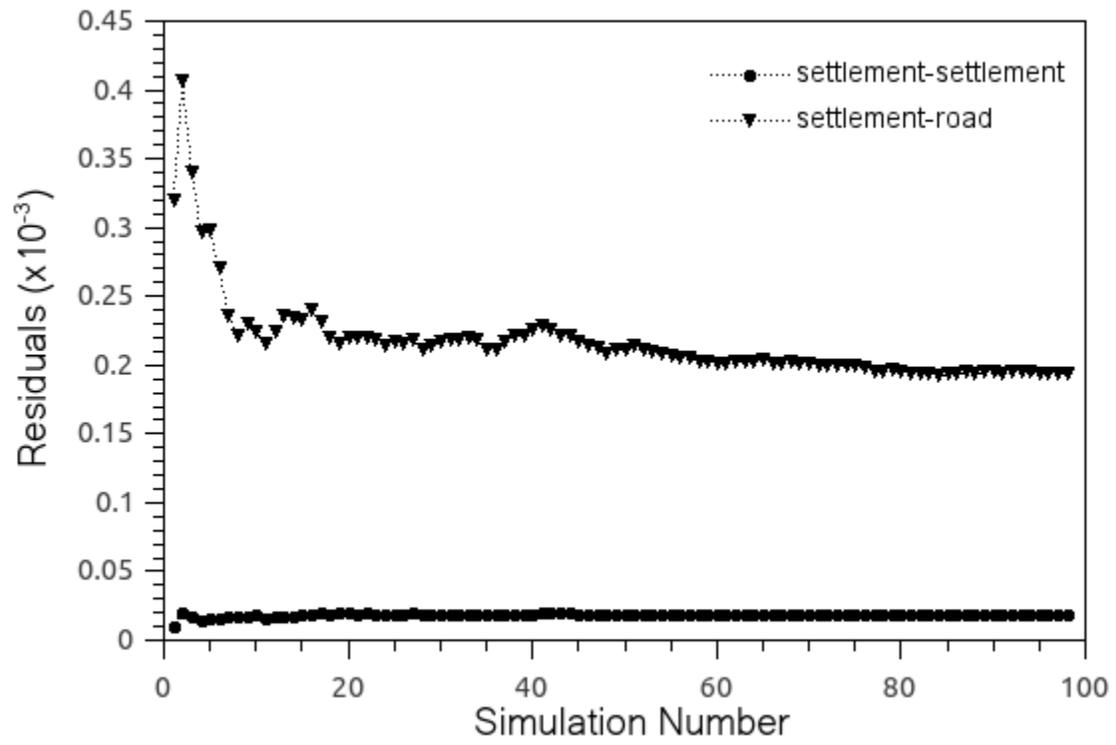